\documentclass[aps,prb,twocolumn,footinbib,floatfix,showkeys,citeautoscript]{revtex4}

\usepackage{amsmath}
\usepackage{amssymb}
\usepackage{amsfonts}
\usepackage{graphicx,graphics}

\newcommand{\ket}[1]{\lvert #1\rangle}
\newcommand{\bra}[1]{\langle#1 \rvert}
\newcommand{\abs}[1]{\lvert #1 \rvert}
\newcommand{\expect}[1]{\langle #1\rangle}

\newcommand{\br}{\mathbf{r}}

\begin{document}

\title{Image charge effects in single-molecule junctions: Breaking of symmetries
  and negative differential resistance in a benzene transistor}
\author{K. Kaasbjerg$^1$}
\email{cosby@fys.ku.dk}
\author{K. Flensberg$^2$}
\affiliation{$^1$Center for Atomic-scale Materials Design (CAMD), 
  Department of Physics, Technical University of Denmark, 
  DK-2800 Kgs. Lyngby, Denmark}
\affiliation{$^2$Niels Bohr Institute and Nano-Science Center, University of Copenhagen,
  Universitetsparken 5, DK-2100 Copenhagen, Denmark}
\date{\today}

\begin{abstract}
  Both experiments and theoretical studies have demonstrated that the
  interaction between the current carrying electrons and the induced
  polarization charge in single-molecule junctions leads to a strong
  renormalization of molecular charging energies. However, the effect on
  electronic excitations and molecular symmetries remain unclear. Using a
  theoretical framework developed for semiconductor nanostructure based
  single-electron transistors (SETs), we demonstrate that the image charge
  interaction breaks the molecular symmetries in a benzene based single-molecule
  transistor operating in the Coulomb blockade regime. This results in the
  appearance of a so-called blocking state, which gives rise to negative
  differential resistance (NDR). We show that the appearance of NDR and its
  magnitude in the symmetry-broken benzene SET depends in a complicated way on
  the interplay between the many-body matrix elements, the lead tunnel coupling
  asymmetry, and the bias polarity. In particular, the current reducing property
  of the blocking state causing the NDR, is shown to vanish under strongly
  asymmetric tunnel couplings, when the molecule is coupled stronger to the
  drain electrode. The calculated IV characteristic may serve as an indicator
  for image charge broken molecular symmetries in experimental situations.
\end{abstract}

\keywords{Single-molecule transistor, Coulomb blockade, junction polarization,
  image charge, rate-equations, benzene, broken symmetry, selection rules, NDR}
\maketitle

\section{Introduction}

Over the past decade the field of single-molecule electronics has experienced
significant progress both experimentally and theoretically. On the experimental
side break-junction and electromigration techniques have become standard ways of
realizing nano-scale junctions with only one molecule bridging the gap between
the source and drain
electrodes~\cite{Park:C60,Natelson:Electromigration,Zant:ReviewJPCM}. With the
coupling between the molecule and the electrodes often being a highly
uncontrollable parameter, the transport through single-molecule junctions spans
different parameter regimes depending on the fabrication
techniques~\cite{Moth:NatureNano}. Three-terminal measurements where the
molecule couples capacitively to a third gate electrode most often end up in the
weak coupling regime where the Coulomb blockaded transport through the molecule
is dominated by sequential tunneling. In these setups, the coupling to the gate
electrode allows control of the alignment between the molecular levels and the
Fermi levels of the leads. The IV characteristics is therefore often summarized
in the so-called charge stability diagram, which maps out the differential
conductance $dI/dV_{\text{sd}}$ as a function of the gate and source-drain
voltage. These diagrams help to understand the excitations intrinsic to the
molecule, as for example the vibrational degrees of
freedom~\cite{Park:C60,Zant:OPV5Vibrations}.

One still unresolved issue is how the junction polarization resulting from the
charging of the molecule influences the molecular states. In, for example,
three-terminal electromigrated nanoscale
junctions~\cite{Park:C60,Natelson:Electromigration}, where the dimensions of the
source and drain electrodes are large compared to the metallic screening length,
this effect can be expected to be significant. So far, both
experiments~\cite{Bjornholm:OPV5,Zant:OPV5Excitations} and theoretical
simulations~\cite{Hedegaard:OPV5,Kaasbjerg:OPV5SET} have demonstrated that the
interaction with the polarization charge---the so-called image charges---leads
to a strong renormalization of the molecular charging energies. In a
charge-stability diagram, this is reflected in addition energies reduced with up
to several electron-volts compared to the value expected from gas phase level
positions of the molecule. On the other hand, the effect on excited states
remain unresolved. A number of experimental studies have indicated image charge
stabilized states close to the electrodes and modified excitation
energies~\cite{Zant:OPV5Excitations,Balestro:NatC60Cotunneling}. Hence, a
thorough investigation of the image charge effect and its influence on model
parameters can help to improve the interpretation of experimental observations.

With the recent interest in the role of molecular degeneracies in the Coulomb
blockade
regime~\cite{Begemann:InterferenceEffects,Cao:InterferenceEffects,Schultz:Degeneracies},
it is of relevance to address the effect on such degeneracies. While splittings
of degenerate levels on the order of the tunnel broadening results in
quasi-degenerate states that still behave as being degenerate, a larger
splitting completely destroys interference effects and the associated signature
in the transport
characteristics~\cite{Begemann:InterferenceEffects,Schultz:Degeneracies}.
Degeneracies also play an important role in Jahn-Teller active molecules that
undergo distortions upon charging. Here, the higher-dimensional adiabatic
potential energy surface of the charged molecule resulting from the coupling to
the Jahn-Teller active vibrations leads to distinct transport
characteristics~\cite{Schultz:BerryPhase,Frederiksen:JahnTellerC60}. Also in
this case can a splitting of the degenerate levels result in a qualitatively
different signature in the stability diagram. For example, a transport signature
characteristic of the pseudo Jahn-Teller effect might result when the size of
the energy splitting matches a multiple of a vibrational
energy~\cite{Leijnse:PseudoJahnTeller}.

In this paper, we present a quantitative study of the impact on molecular
degeneracies by studying the image charge effect in a benzene-based
single-molecule junction. Despite the presence of the Jahn-Teller effect in the
benzene molecule~\cite{Allen:C6H6JahnTeller}, we will focus on the splitting of
the electronic states due to image charge effects and neglect any coupling to
vibrations.

Recent theoretical studies have already touched upon the image charge effect and
other symmetry-breaking agents in a benzene SET, however, without taking into
account the full interaction between the molecule and the image
charges~\cite{Begemann:InterferenceEffects,Paaske:Broken}. Here, the purpose is
twofold. First, we present a general framework that has previously been used to
account for image charge effects in semiconductor
nanostructures~\cite{Maksym:Screening} and discuss its applicability for
single-molecule junctions operating in the Coulomb blockade regime. Second, the
implication on the degeneracies of the benzene molecule and the consequence for
the low-bias transport characteristics are investigated. We find that the image
charge effect indeed breaks the high symmetry of the molecule and leads to a
large splitting ($\sim40-80$ meV) of the degenerate ground-state of the
singly-charged molecule.

Due to a breakdown of the transport selection rules that apply in the isolated
molecule, the symmetry-split excited state of the charged molecule is turned
into a so-called blocking state. As a consequence, an NDR feature appears at a
bias corresponding to the level splitting. The stability of the NDR feature with
respect to an asymmetry in the lead couplings and the coupling site on the
molecule is analyzed. Similar NDR features caused by radiative relaxation to a
blocking state~\cite{Hettler:CurrentCollapse} and an interference-induced
blocking state~\cite{Begemann:SymmetryFingerprints,Begemann:InterferenceEffects}
have been reported previously in the literature for the benzene SET. Contrary to
these cases, the NDR feature observed here is exclusively caused by the broken
symmetry of the molecule and needs none of these additional effects to occur.

The paper is organized as follows. Sec.~\ref{sec:hamiltonian} presents the
theoretical approach used to describe the image charge effect in this work. This
involves a generalized Hamiltonian for the molecule with extra terms originating
from the interaction with the image charges. In Sec.~\ref{sec:benzene_set}, the
benzene SET and the semi-empirical Pariser-Parr-Pople Hamiltonian is introduced.
Furthermore, an analysis of the broken symmetry and the accompanying splitting
of the degenerate molecular states together with a condition for the occurrence
of NDR is given. The resulting low-bias IV characteristics are presented and
analyzed in Sec.~\ref{sec:iv}. Finally, Sec.~\ref{sec:conclusion} summarizes our
findings and points to other situations where the image charge effect may be
important in single-molecule junctions.
\begin{figure*}[!t]
  \centering
  \includegraphics[width=0.95\linewidth]{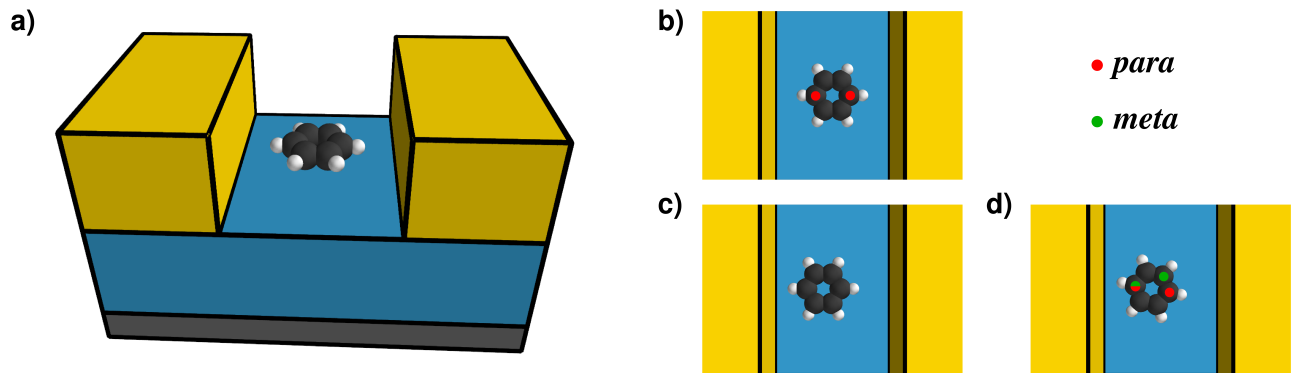}
  \caption{(Color online) Schematic illustration of a benzene single-electron
    transistor. a) Side view of the nanojunction with the molecule positioned on
    the gate dielectric between the source and drain electrodes. The capacitive
    coupling between the gate electrode (gray) at the bottom and the molecule,
    provides control over the energy levels of the molecule. b), c) and d) Top
    view of the junction with the molecule positioned in different
    configuration. The red and green dots indicate the coupling sites for
    coupling in the \emph{para} and \emph{meta} configuration, respectively. b)
    Symmetric setup with the molecule positioned in the middle of the gap
    between the electrodes. c) Asymmetric setup with the molecule positioned
    closer to the left electrode. d) Same as in c) but with an additional
    rotation of the molecule. Due to the interaction with the image charges of
    the junction, the rotation of the molecule breaks all the symmetries of the
    molecular Hamiltonian and results in qualitatively different
    IV-characteristics and NDR.}
\label{fig:benzene_set}
\end{figure*}

\section{Junction Hamiltonian and the current}
\label{sec:hamiltonian}

In the following section, we present a general framework for the description of
single-molecule junctions operating in the Coulomb blockade regime, i.e. with a
weak tunnel coupling between the molecule and the leads. While different
approaches to the calculation of the current to different orders in the leads
couplings have been given elsewhere~\cite{Leijnse:Kinetic}, the present study
focuses on the interaction with the image charges in the junction environment.
Figure~\ref{fig:benzene_set}a illustrates an idealized setup for such a
three-terminal junction with the molecule lying on a gate oxide between the
source and drain electrodes. With the molecule playing the role of the quantum
dot, this type of single-molecule junctions is very similar to multi-terminal
semiconductor-based quantum dot structures, where the theoretical description of
image charges and their influence on the dot states is
well-established~\cite{Maksym:Screening}. The theoretical framework is based on
a macroscopic description of the metallic electrodes and the surrounding
dielectric, while the quantum dot(s) is described quantum mechanically.

It is here illustrated how the same formalism can be applied to single-molecule
junction. We briefly summarize the most important aspects of the approach and
explain in detail the appearance of new terms in the molecular Hamiltonian that
result from the image charge effect. The validity of a macroscopic electrostatic
description for single-molecule SETs is also discussed.

\subsection{Hamiltonian}

A general Hamiltonian for a single-molecule junction including the image charge
effect, can be written as the sum of the following terms,
\begin{equation}
  \label{eq:H}
  H = H_{\text{mol}} + H_{\text{T}} + H_{\text{leads}} + H_{\text{mol-env}} + H_{\text{env}} .
\end{equation}
Here, $H_{\text{mol}}$ denotes the Hamiltonian of the molecule expressed in a
basis of atomic orbitals ${\phi_i}$ with corresponding creation and annihilation
operators $\{c^{\dagger}_{i\sigma}, c^{\phantom\dagger}_{i\sigma}\}$. In the
present work, this basis will be the atomic p$_z$ orbitals of the carbon atoms
in the benzene molecule. The next two terms account for the tunnel couplings to
the lead (labeled by $\alpha$) electrons,
\begin{equation}
  \label{eq:H_t}
  H_{\text{T}} + H_{\text{leads}} = 
      \sum_{k\sigma\alpha} t^{\alpha}_{k\sigma}c^{\dagger}_{k\sigma\alpha}c^{\phantom\dagger}_{\alpha\sigma} + \text{h.c.} +
      \sum_{k\sigma\alpha} \varepsilon^{\alpha}_{k\sigma}
          c^{\dagger}_{k\sigma\alpha}c^{\phantom\dagger}_{k\sigma\alpha} .
\end{equation}
For simplicity, it is here assumed that only one orbital $\phi_{\alpha}$ on the
molecule couples to each of the left ($\alpha=L$) and right ($\alpha=R$) leads.

The remaining two terms in Eq.~\eqref{eq:H} account for the electrostatic energy
arising from the interaction between the charge of the molecule and
junction. This includes the interaction with the induced polarization charge in
the environment due to charging of the molecule (the image charges), the energy
of the molecule in the potential from voltages $V_i$ applied to the electrodes,
and the electrostatic energy of the capacitively coupled electrodes of the
junction. For now, these two terms are most conveniently expressed together with
the molecular Hamiltonian $H_{\text{mol}}$ in real-space and first-quantization
as~\cite{Maksym:Screening}
\begin{widetext}
\begin{equation}
  \label{eq:H_molenv}
  H_{\text{mol}} + H_{\text{mol-env}} + H_{\text{env}} =
      \frac{1}{2m} \sum_{n=1}^N \nabla_n^2 - 
      e \sum_{n=1}^N V(\br_n) +
      \frac{e^2}{2} \sum_{n=1}^N \sum_{n'\neq n} G(\br_n, \br_{n'}) +
      \sum_{ij} V_i C_{ij} V_j ,
\end{equation}
\end{widetext}
where $m$ is the electron mass, $N$ is the number of electrons $N$ on the
molecule, $G$ denotes the electrostatic Green's function of the junction,
$C_{ij}$ is the capacitance matrix of the junction coupling the charge on the
different electrodes. The single-particle potential $V$ is given by
\begin{equation}
  \label{eq:V}
  V(\br) = - \frac{e}{2} \widetilde{G}(\br, \br) + 
             \int d\br' G(\br, \br') \rho_{\text{ion}}(\br') +
             \sum_i \alpha_i(\br) V_i ,
\end{equation}
where $\widetilde{G}$ is the part of the Green's function that accounts for the
induced potential due to an electron in $\br$ (see below), $\rho_{\text{ion}}$
is the charge distribution of ionic cores of the molecule (often described by a
pseudo-potential in \emph{ab-initio} approaches), and $\alpha_i$ is an electrode
specific function, which gives the spatial profile of the junction potential
with a unit voltage applied to the $i$'th electrode and in the absence of the
molecule. The $\alpha$-functions are solutions to Laplace equation
\begin{equation}
  \label{eq:Laplace}
  - \nabla \cdot \left[ \epsilon_0 \varepsilon_{\text{r}}(\br) \nabla \alpha_i(\br) \right]
  = 0
\end{equation}
with boundary conditions $V_j = \delta_{ij}$ at the electrodes,
$\varepsilon_{\text{r}}$ is the spatially varying dielectric constant of the
junction, and $\epsilon_0$ is the vacuum permittivity.

Apart from the capacitive energy in last term of Eq.~\eqref{eq:H_molenv} and the
contribution from the applied voltages in the last term of Eq.~\eqref{eq:V}, the
Hamiltonian in Eq.~\eqref{eq:H_molenv} bears close resemblance to the usual
many-body Hamiltonian for interacting electrons moving in an external potential.
However, the electrostatic Green's function plays an important role in this
modified Hamiltonian. It is seen to have replaced the Coulomb interaction (third
and second terms of Eqs.~\eqref{eq:H_molenv} and~\eqref{eq:V}, respectively) and
results in an additional single-particle term (first term of Eq.~\eqref{eq:V}).

The electrostatic Green's function solves Poisson's equation with a
$\delta$-function source term,
\begin{equation}
  \label{eq:PoissonGF}
  - \nabla \cdot \left[ \epsilon_0 \varepsilon_{\text{r}}(\br) \nabla G(\br, \br') \right] 
  = \delta(\br - \br')
\end{equation}
with Dirichlet boundary conditions $V_i = 0$. For a given junction
geometry the Green's function gives the potential in $\br$ due to a unit charge
in $\br'$. When both $\br$ and $\br'$ belong to the vacuum region of the
junction where the molecule resides, the Green's function can be written as a
sum of the direct (unscreened) Coulomb interaction plus a contribution
$\widetilde{G}$ from the image charges (screening),
\begin{equation}
  G(\br, \br') = \frac{1}{\vert \br - \br' \vert} + \widetilde{G}(\br, \br'). 
\end{equation}
The replacement of the Coulomb interaction with the electrostatic Green's
function in Eqs.~\eqref{eq:H_molenv} and~\eqref{eq:V} therefore corresponds to a
screening of the Coulomb interaction by the polarization response of the
junction (the image charges). The additional single-particle term in
Eq.~\eqref{eq:V} has the form of an electronic self-interaction given by the
$\widetilde{G}$-part of the Green's function. This is the energy of the electron
in its own induced potential, i.e. the interaction between the electron and its
own image charge. For the standard example of a point charge positioned at a
distance $z$ from an infinite conduction surface this term reduces to the
classical $-1/4z$-energy well known from classical
electrostatics~\cite{Griffiths}.

In the present work, the electrostatic Green's function is obtained for the
simplified junction geometry discussed in App.~\ref{app:SimpleJunction}. This
allows for an analytical solution of Poisson's equation in
Eq.~\eqref{eq:PoissonGF}. It has been verified that this gives a very good
description of the potential in the realistic junction in
Fig.~\ref{fig:benzene_set}~\cite{Kaasbjerg:OPV5SET}. The spatial profiles of the
source-drain and gate voltages follows from the $\alpha$-functions which are
solutions to Laplace equation in Eq.~\eqref{eq:Laplace}. For simplicity, we
approximate these by simple linear functions. The coupling to the gate electrode
is set to unity, i.e. $\alpha_{\text{gate}} = 1$. In realistic descriptions of
the gate potential and in experimental realizations of nanoscale junctions this
number is often rather low ($\sim
0.1-0.2$)~\cite{Kaasbjerg:OPV5SET,Johnson:GateCoupling,Zant:OPV5Excitations,Bjornholm:OPV5}
and may vary over the spatial extend of the molecule~\cite{Kaasbjerg:OPV5SET}.
However, under the assumption that the gate potential is constant on the
molecule, the gate-coupling parameter only serves as a scaling factor for the
shift of the molecular energy levels. The source-drain voltage is modeled by a
linear ramp between the two electrodes with the voltage applied symmetrically to
the left and right electrode,
\begin{equation}
  \label{eq:bias_voltage}
  V(z) = \frac{V_{\text{sd}}}{2} - V_{\text{sd}} \frac{z}{L} ,
\end{equation}
where $z=0$ corresponds to the position of the left electrode and $L$ is the
electrode spacing. With the chemical potentials of the leads given by
\begin{equation}
  \label{eq:chemical_potentials}
  \mu_L = E_f - eV_{\text{sd}}/2 
  \quad \text{and} \quad
  \mu_R = E_f + eV_{\text{sd}}/2 ,
\end{equation}
where $E_f$ is the equilibrium Fermi level, the positive current direction is
from left to right. Within a microscopic description of both the leads and the
molecule, the alignment between the lead Fermi levels and the molecular levels
follows directly. However, for the simplified description of the leads adopted
here, the alignment is treated as a parameter, and the equilibrium Fermi levels
are chosen to reside in the middle of the gap of the benzene molecule. The
level alignment is illustrated schematically in Fig.~\ref{fig:overview}b for a
situation with voltages applied to the electrodes.

\subsection{Validity of an electrostatic description}

The classical electrostatic treatment of the junction environment in the
Hamiltonian in Eq.~\eqref{eq:H} relies on certain assumptions. First of all, the
dimensions of the electrodes and gate dielectric need to be larger than the
screening lengths in the respective materials in order for a classical
description to be valid. Secondly, the classical treatment of the environment
assumes that there is no significant overlap between the quantum mechanical
region and the environment. In the Coulomb blockade regime, this seems to be a
fair assumption. Last, since the environment is described with electrostatics,
the time scale of the electrons on the molecule needs to be slower than the
response times of the metallic electrodes and gate dielectric. This requires the
hopping integrals $t$ on the molecule to be smaller than typical values for the
plasmon and phonon energies. With molecular level spacing being on the order of
eV for conjugated molecules, this is not necessarily the case. Nevertheless, the
current carrying electrons have residence times on the molecule much longer than
the response time of the environment. So the question is not whether or not the
image charge effect should be included, but rather if it should be treated as an
instantaneous interaction as in Eq.~\eqref{eq:H_molenv} or at the level of
mean-field theory where the environment only sees the mean occupation of the
electrons on the molecule. In the following we pursue the first direction. We
have verified that a mean-field treatment at the level of Hartree-Fock does not
change the main conclusions of the present work.

\subsection{Current}
\label{sec:rate_equations}

We shall here focus on the weak-coupling regime where the occupation
probabilities of the molecular states and the current can be obtained within a
master equation approach. In the case of degeneracies between the molecular
states, a master equation for the density matrix that retains the coherence
between the degenerate states must be
considered~\cite{Begemann:SymmetryFingerprints,Schultz:Degeneracies}. However,
it turns out that in the present study such degeneracies are broken by the
interaction with the image charges in the junction environment, which leaves the
simpler rate-equation approach valid, so that only the diagonal elements of the
reduced density matrix need to be retained~\cite{Flensberg,Ralph:RateEqs}.

To lowest order in the coupling between the molecule and the leads in
Eq.~\eqref{eq:H_t}, the transition rate between the $i$'th $N$-electron state
and the $j$'th $N+1$-electron state due to tunneling from lead $\alpha$ is given
by Fermi's golden rule
\begin{equation}
  \label{eq:Gamma_in}
  \Gamma_{N+1,j \atop N,i}^{\alpha\sigma} = 
    \frac{2\pi}{\hbar} \gamma^{\alpha\sigma}_{j,i}(N) f_{\alpha}(E_{ij}) ,
\end{equation}
where $f_{\alpha}$ is the Fermi distribution for the lead electrons,
$E_{ij}=E^{N+1}_j-E^N_i$ corresponding to the molecular ionization energies and
electron affinities, and
\begin{equation}
  \label{eq:gamma}
  \gamma_{j,i}^{\alpha\sigma}(N) = \rho_{\alpha} t_{\alpha}^2
  \left\vert \langle N+1,j \vert  c^{\dagger}_{\alpha\sigma} 
      \vert N,i \rangle \right\vert^2   
\end{equation}
is the product of the lead density of states $\rho_{\alpha}$, the tunnel
coupling $t_{\alpha}$, and the transition matrix element between the two
molecular states involved in the addition of an electron to the molecule from
lead $\alpha$. For the opposite process where an electron tunnels from the
molecule to the drain electrode, the transition rate is given by
\begin{equation}
  \label{eq:Gamma_out}
  \Gamma_{N-1,j \atop N,i}^{\alpha\sigma} = 
    \frac{2\pi}{\hbar} \gamma_{i,j}^{\alpha\sigma}(N-1) 
    \Big(1 - f_{\alpha}(E_{ij}) \Big)
\end{equation}
where the fermi factor ensures that there is an empty state in the lead. We here
restrict our discussion to the case of identical left and right leads, modeled
by normal metals with a constant density of states. The presence of a fast (on
the scale of the time between tunneling events) energy relaxation mechanism in
the leads justifies the equilibrium description of the lead electrons in
Eqs.~\eqref{eq:Gamma_in} and~\eqref{eq:Gamma_out}.

The rate-equations for the occupations of the molecular states now reads
\begin{align}
  \dot{P}_{N,i} = \sum_{\alpha,j} \bigg[
    & - P_{N,i} 
      \left( \Gamma^{\alpha}_{N+1,j \atop N,i} + 
        \Gamma^{\alpha}_{N-1,j \atop N,i} \right)  \nonumber \\
    & + P_{N+1,j}\Gamma^{\alpha}_{N,i \atop N+1,j} +
        P_{N-1,j}\Gamma^{\alpha}_{N,i \atop N-1,j} \bigg] .
\end{align}
Together with the normalization condition $\sum_{N,i}P_{N,i}=1$ the
rate-equations can be solved in steady-state, i.e. $\dot{P}_{N,i}=0$, for the
occupation probabilities $P_{N,i}$. From the steady-state occupations, the
current through the molecule follows by evaluating the total rate of electrons
from lead $\alpha$
\begin{equation}
  \label{eq:current}
  I_{\alpha} = \mp e \sum_{N, ij} 
      \left( 
        P_{N, i} \Gamma^{\alpha}_{N+1,j \atop N,i} - 
        P_{N, i} \Gamma^{\alpha}_{N-1,j \atop N,i}
      \right) 
\end{equation}
for $\alpha = L/R$, respectively.

The main objective is here to demonstrate how the inclusion of the image charge
effect affects the molecular states and thereby also the IV characteristic of
the molecular junction. For this purpose, the energies and states appearing in
Eqs.~\eqref{eq:Gamma_in},~\eqref{eq:gamma}, and~\eqref{eq:Gamma_out} must be
calculated on the basis of the junction Hamiltonian in Eq.~\eqref{eq:H_molenv}
which fully accounts for the interaction with the image charges. The extend to
which specific molecular properties are affected is highly dependent on the
molecule of interest and its configuration in the nanojunction. A general
discussion is therefore not possible, why the rest of the paper seeks to address
the importance of the image charge effect in the often studied benzene
SET~\cite{Hettler:CurrentCollapse,Begemann:SymmetryFingerprints,Begemann:InterferenceEffects}.

\section{Benzene SET}
\label{sec:benzene_set}

In the remaining of the paper we consider the transport through the benzene SET
illustrated in Fig.~\ref{fig:benzene_set}a. It has the molecule lying flat on a
gate oxide between the source and drain electrodes. The dielectric constant of
the gate oxide is set to $\varepsilon_{\text{r}} = 10$ corresponding to the
high-$\kappa$ dielectric Al$_2$O$_3$.

In experimental situations both the position of the molecule and the electrode
spacing are highly uncontrollable parameters. We therefore consider different
setups where the position and orientation of the molecule with respect to the
electrodes are varied. In the symmetric setup shown in
Fig.~\ref{fig:benzene_set}b, the molecule is placed in the middle of the
junction with the two end atoms facing the electrodes~\cite{note1}. In this
reference setup, the distance between the hydrogen atoms facing the electrodes
at the ends of the molecule and the electrostatic boundaries of the electrodes
is set to 1.2~{\AA}. With the so-called image charge plane lying $\sim
1.0$~{\AA} outside the outermost atomic layer of a surface~\cite{Needs:Image},
this corresponds to a weak bond between the surface atoms of the electrodes and
the benzene molecule (see also App.~\ref{app:SimpleJunction}). To account for
the longer range of the p$_z$-orbitals on the carbon atoms, the distance to the
gate dielectric is chosen slightly larger to 2 {\AA}. In the following, all
changes will be with respect to the above described reference setup.

In order to model experimentally more relevant situations, we also consider the
following two setups and combinations hereof. In the first, the molecule is
rotated by an angle $\theta$ with respect to its symmetric setup in
Fig.~\ref{fig:benzene_set}b around its six-fold rotational symmetry axis
perpendicular to the plane of the molecule. In the second, an asymmetric setup
where the distance to one of the electrodes is increased to twice the distance
in the reference setup is considered. This leads to a smaller image charge
interaction with the most distant electrode. While the symmetric setup
corresponds to electrode couplings in the \emph{para} configuration (marked by
red dots in Fig.~\ref{fig:benzene_set}), coupling also in the \emph{meta}
configuration (green dots in Fig.~\ref{fig:benzene_set}) is likely to occur
in the asymmetric setup where the relative difference in the distance between
the distant electrode and the two coupling sites is smaller. The combination of
the above described situations is shown in Fig.~\ref{fig:benzene_set}d where the
molecule is placed in a rotated position closer to the left electrode.

\subsection{Pariser-Parr-Pople Hamiltonian}

In conjugated molecules the sp$^2$ hybridization of the carbon atom results in
an energy separation between the bonding $\sigma$-orbitals and the higher lying
$\pi$-orbitals which have mainly p$_z$ character. Quantitative predictions of
the low energy excitations can therefore be obtained with the simple
Pariser-Parr-Pople~\cite{PariserParrI,PariserParrII,Pople} (PPP) description.
The PPP Hamiltonian which includes only the $\pi$-electron system is given by
\begin{align}
  \label{eq:H_PPP}
  H_{\text{mol}} = & \sum_{i\sigma} \varepsilon_{i\sigma} \hat{n}_{i\sigma}
      - \sum_{\expect{ij} \sigma} t_{ij} 
         \left[ 
           c^{\dagger}_{i\sigma}c^{\phantom\dagger}_{j\sigma} + \mathrm{h.c.}
         \right]  \nonumber \\
    & + \frac{1}{2} \sum_{i\neq j} V_{ij} (\hat{n}_i-Z_i) (\hat{n}_j-Z_j)
      + \sum_i U_i \hat{n}_{i\uparrow}\hat{n}_{i\downarrow} ,
\end{align}
where $\hat{n}_i = \hat{n}_{i\uparrow} + \hat{n}_{i\downarrow}$ and $Z_i=1$ are
the occupation and valence, respectively, of the p$_z$-orbital at the carbon
site $i$ of the molecule. Apart from hopping between the p$_z$ orbitals on
neighboring sites in the second term, the PPP description also includes onsite
and long-ranged Coulomb interactions between the $\pi$-electrons. Due to the
effective $\pi$-electron description, the parameters $U$ and $V$ cannot be
identified with the usual matrix elements of the Coulomb interaction since
screening effects from the $\sigma$-electrons reduce the Coulomb interaction.
Here, the Ohno parametrization~\cite{Ohno} is used for the long-ranged Coulomb
interactions,
\begin{equation}
  \label{eq:Ohno}
  V_{ij} = \frac{U}{\sqrt{1 + (\alpha r_{ij})^2}},
\end{equation}
where $\alpha=U/(14.397\,\mathrm{eV}\,\mathrm{\AA})$ and $r_{ij}$ is the
distance between the carbon sites $i$ and $j$. It ensures that the bare Coulomb
interaction is recovered for large distances, i.e. $V_{ij}\rightarrow 1/r$ as
$r_{ij}\rightarrow \infty$, while for shorter distances the (screened) onsite
Coulomb interaction $U$ is approached. The remaining parameters of the PPP
Hamiltonian in Eq.~\eqref{eq:H_PPP} have been chosen as $\varepsilon_i =
0\;\mathrm{eV}$, $t=2.539\;\mathrm{eV}$ and $U=10.06\;\mathrm{eV}$. This set of
parameters have been fitted to experimental excitation energies and hence gives
a quantitative description of the excitations in the benzene
molecule~\cite{Barford:Benzene}. We note that in previous studies of Coulomb
blockade transport through the benzene
molecule~\cite{Hettler:CurrentCollapse,Begemann:SymmetryFingerprints,Begemann:InterferenceEffects},
different sets of parameters have been used. The excitation spectrum of the
isolated benzene molecule reported here, might therefore differ slightly from
these other works. It should be stressed that despite the large value of the
onsite Coulomb interaction $U$, the inclusion of long-ranged Coulomb
interactions $V_{ij}$ in the PPP description results in weakly correlated states
that to a good approximation can be described with Hartree-Fock
theory~\cite{Kaasbjerg:Benchmark}.

The PPP Hamiltonian in Eq.~\eqref{eq:H_PPP} describes only the isolated
molecule. In order to include the interaction with the junction environment, the
terms in the Hamiltonian in Eq.~\eqref{eq:H_molenv} involving the
$\widetilde{G}$ part of the Green's function and the applied voltages $V_i$ must
be taken care of separately. The conversion of these terms from the real-space
representation in Eq.~\eqref{eq:H_molenv} to the atomic p$_z$-orbital basis of
the PPP Hamiltonian in Eq.~\eqref{eq:H_PPP} is given in
App.~\ref{app:MatrixElements}.  This leads to the following additional terms
\begin{align}
  \label{eq:H_PPP_molenv}
  H_{\text{mol-env}} + H_{\text{env}} = & \sum_{i\sigma} 
      \left[ 
          \widetilde{V}_i^{\text{ion}} + \frac{1}{2}\widetilde{V}_{ii} 
          + V_i^{\text{ext}}
      \right] \hat{n}_{i\sigma} \nonumber \\
  & + \frac{1}{2} \sum_{i \ne j} \widetilde{V}_{ij}  \hat{n}_{i} \hat{n}_{j}
    + \sum_i \widetilde{V}_{ii} \hat{n}_{i\uparrow} \hat{n}_{i\downarrow} ,
\end{align}
where $\widetilde{V}_{i}^{\text{ion}}$, $V_{i}^{\text{ext}}$, and
$\widetilde{V}_{ij}$ are the matrix elements of the induced junction potential
$\widetilde{V}_{\text{ion}}$ due to the ionic cores, the matrix element of the
external potential $V_{\text{ext}}=\sum_i \alpha_i V_i$ from the applied
voltages $V_i$, and the two-particle matrix elements of $\widetilde{G}$ in the
basis of the atomic p$_z$-orbitals, respectively. In order to adapt to the level
of complexity of the PPP Hamiltonian, only the diagonal matrix elements of the
terms in Eq.~\eqref{eq:H_PPP_molenv} are retained here. This should not affect
the main conclusions of the paper.
\begin{figure}[!t]
  \centering
  \includegraphics[width=0.99\linewidth]{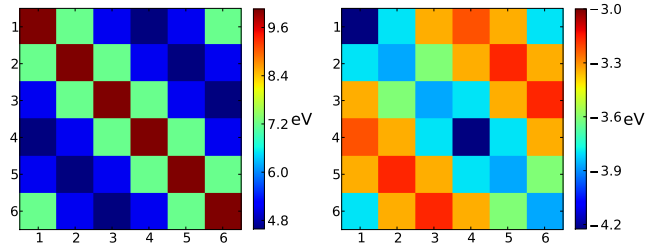}
  \caption{(Color online) Matrix elements (in eV) of the Coulomb interaction,
    given by the Ohno parametrization in Eq.~\eqref{eq:Ohno} (left), and the
    image charge interaction $\widetilde{V}_{ij}$ in Eq.~\eqref{eq:H_PPP_molenv}
    (right). The numbers on the axes denote the site indices of the carbon atoms
    in the benzene molecule with site 1 and 4 corresponding to the carbon sites
    marked with red dots in Fig.~\ref{fig:benzene_set}b. The large value of the
    matrix elements of the image charge interaction results in a significant
    renormalization of the benzene Hamiltonian and a breaking of its high
    six-fold rotational symmetry.}
\label{fig:matrix_elements}
\end{figure}

As is evident from Eq.~\eqref{eq:H_PPP_molenv}, the interaction with the image
charges leads to a renormalization of both the onsite energies and the Coulomb
interaction. The two first terms inside the square brackets stem from the image
charge of the ions and the image charge of the electron itself. The third term
inside the brackets is the shift in the onsite energies due to the applied
voltages. The two last terms in the last line account for the interaction with
the image charges from all the other electrons. These terms have the usual form
of the Coulomb interaction and thus correspond to a renormalization of the
interactions of the bare molecule in Eq.~\eqref{eq:Ohno}. 

The matrix elements of the Coulomb interaction in Eq.~\eqref{eq:Ohno} and the
image charge interaction $\widetilde{V}_{ij}$ (see App.~\ref{app:MatrixElements}
for details) are shown in Fig.~\ref{fig:matrix_elements} for the symmetric
setup. As can be seen, the matrix elements of the image charge interaction is on
the order of several electron-volts and therefore result in a strong
renormalization of the Coulomb interactions on the molecule. With the large
renormalization of both the onsite energies and the Coulomb interaction in the
molecular Hamiltonian in Eq.~\eqref{eq:H_PPP}, its is not surprising that the
image charge effect has a considerable impact on the molecular states, their
energies and symmetry.

In the following the many-body Hamiltonian given by the sum of the contributions
in Eqs.~\eqref{eq:H_PPP} and~\eqref{eq:H_PPP_molenv} is diagonalized directly in
the Fock space of many-body states. Since the Hamiltonian commutes with the
number operator $\hat{N}=\sum_{i\sigma} \hat{n}_{i\sigma}$ and the
$z$-projection $S_z$ of the total spin, each of the $(N, S_z)$-subblocks of the
Fock space are diagonalized separately. For the neutral $N=6$ state of the
molecule, the dimension of this subblock is $400\times 400$ implying that the
Hamiltonian can be diagonalized with standard diagonalization routines. This
yields the many-body states
\begin{equation}
  \label{eq:state}
  \ket{N,i}=\sum_n c_n \ket{\phi_n^N}
\end{equation}
and their corresponding energies $E_i^N$, where $\{\ket{\phi_n^N}\}$ denotes the
possible $N$-electron configurations with spin $S_z$ and $c_n$ are the expansion
coefficients. For example, in the singlet ground-state of the neutral isolated
molecule where the Hamiltonian is given by Eq.~\eqref{eq:H_PPP}, the
configuration
$\ket{\phi_n^{N=6}}=\ket{\uparrow\downarrow\uparrow\downarrow\uparrow\downarrow}$
with the spins aligned oppositely on neighboring sites is the one with the
largest weight.

\begin{figure}[!t]
  \centering
  \includegraphics[width=0.98\linewidth]{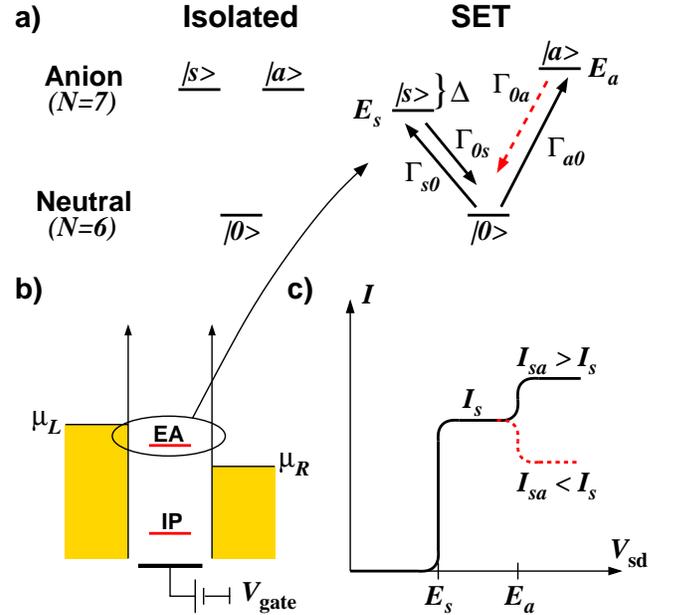}
  \caption{(Color online) Schematic overview of the states and transport
    characteristics of the benzene SET. a) States of the benzene molecule
    important for the low-bias transport through the molecule. In the SET
    environment (right), the image charge effect lifts the degeneracy of the
    anionic ground-state of the isolated molecule (left) producing a splitting
    $\Delta$ between the two states labeled $\ket{s}$ and $\ket{a}$ (see text).
    The transition rates $\Gamma$ between the states are indicated with arrows.
    The transition marked with the (red) dashed arrow is responsible for the
    occurrence of NDR. b) Alignment between the molecular levels and the the
    Fermi levels of the leads in an out-of-equilibrium situation. c) Current as
    a function of source-drain voltage. With state $\ket{a}$ being a so-called
    blocking state with a slow exit rate illustrated by the dashed red arrow
    in a), it is likely to produce an NDR feature when introduced in the bias
    window.}
\label{fig:overview}
\end{figure}

In the following, we will focus on transport through the benzene molecule at
positive gate voltages. This corresponds to the situation illustrated in
Fig.~\ref{fig:overview}b where the affinity level is located in the bias window.
Due to the electron-hole symmetry of the PPP Hamiltonian, the transport through
the positively charged cation will be identical. The states relevant for the
transport at positive gate voltages are illustrated schematically in
Fig.~\ref{fig:overview}a. This includes the ground-state of the neutral molecule
($N=6$) and the degenerate ground-state of the singly-charged anion ($N=7$)
which is split up by the image charge effect as illustrated in the right plot.
The next excited states lie more than $2$ eV above these states and will hence
not be active under moderate source-drain biases.

\subsection{Addition energy and breaking of degeneracies}

The most apparent effect of the interaction with the image charges is a strong
renormalization of the molecular charging energies which has been observed both
experimentally~\cite{Bjornholm:OPV5,Zant:OPV5Excitations} and
theoretically~\cite{Kaasbjerg:OPV5SET,Hedegaard:OPV5}. This results in a large
reduction of the addition energy, given by the difference between the ionization
potential (IP) and electron affinity (EA),
\begin{equation}
  \label{eq:addition_energy}
  E_{\text{add}} = \text{IP} - \text{EA} = E_0^{N+1} + E_0^{N-1} - 2 E_0^{N}, 
\end{equation}
compared to its value for the isolated molecule. As shown in the first column of
Tab.~\ref{tab:energy_levels}, the addition energy of the benzene molecule is
reduced by up to $\sim 3.6$ eV in setups similar to those in
Fig.~\ref{fig:benzene_set}. This is of the same size as the reductions found
with \emph{ab-initio} descriptions of the benzene molecule in similar
environments~\cite{Louie:Benzene,Stokbro:BenzeneSET}. Due to the smaller image
charge effect from the most distant electrode in the asymmetric setup, the
reduction of the addition energy is smaller in this case.  Similar reductions of
the molecular HOMO-LUMO gaps are expected to occur in single-molecule junctions
with a stronger coupling to the leads where coherent transport
dominate~\cite{Thygesen:Renormalization}.

The large renormalization of the molecular energy levels is also manifested in
the charge distribution on the molecule where the attractive nature of the
oppositely charged image charges polarizes the molecule. The magnitude of this
effect can be quantified in terms of the site occupations on the benzene ring
which follow from the expectation value
$\expect{\hat{n}_i}=\bra{N}\hat{n}_i\ket{N}$. In the symmetric setup in
Fig.~\ref{fig:benzene_set}b, we find that $\sim 0.3$ of the added electron in
the anion resides on each of the end atoms closest to the electrodes while only
$\sim 0.1$ resides on each of the four center atoms. For the exact
diagonalization of the Hamiltonian given here, this charge rearrangement
corresponds to a change in the weights $c_n$ of the different $N$-particle
configurations in the many-body state compared to the isolated molecule.

In transport measurements, the addition energy can be inferred from the height
of the Coulomb diamonds in the charge stability diagram~\cite{Zant:Single}.
However, even when taking into account the image charge effect, the addition
energy of the benzene molecule is large compared to experimentally accessible
source-drain voltages, and hence the observation of full Coulomb diamonds for
such a small molecule seems unlikely. We will therefore focus on the IV
characteristics at lower biases, where the image charge effect leaves its
fingerprint in the form of an additional molecular level that results from a
broken symmetry in the molecule.

The PPP Hamiltonian for the isolated benzene molecule in Eq.~\eqref{eq:H_PPP}
belongs to the $D_{6h}$ point group. The symmetry and degeneracies of the
different charge states of the PPP Hamiltonian for benzene have been considered
in detail in Ref.~\onlinecite{Begemann:InterferenceEffects}. The symmetry of the
ground-states for the neutral molecule and the anion is $A_{1g}$ and $E_{2u}$,
respectively, with the latter having a twofold orbital degeneracy on top of its
spin degeneracy. The states are illustrated schematically in
Fig.~\ref{fig:overview}a.
\begin{table}[!t]
\begin{ruledtabular}
\begin{tabular}{lccccc}
                  & $E_{\text{add}}$  & $\Delta$ & Symmetry  \\  
\hline                                                           %
Isolated          & 11.38             &    --    &  $D_{6h}$ \\  
SET (symmetric)   &  7.77             &    72    &  $D_{2h}$ \\  
SET (asymmetric)  &  8.01             &    43    &  $C_{2v}$ \\  
SET (rotated)     &  7.73             &    76    &     --    \\  
\end{tabular}
\end{ruledtabular}
\caption{Addition energy (in eV) and splitting $\Delta$ (in meV) of the 
  degenerate ground-state of the negatively charged anion ($N=7$) in 
  different situations. The symmetric and antisymmetric configurations
  correspond to the setups illustrated in Figs.~\ref{fig:benzene_set}b 
  and c, respectively. In the rotated setup, the molecule has been 
  rotated by an angle $\theta=\pi/6$ around its six-fold rotational symmetry 
  axis. The reduction of $\sim 3.6$ eV for the addition energy in the 
  symmetric and rotated setups is of the same size as the reductions found with
  \emph{ab-initio} GW/DFT calculations for a benzene molecule on a graphite 
  surface~\cite{Louie:Benzene} and in a SET 
  environment~\cite{Stokbro:BenzeneSET}. The smaller reduction observed for the 
  addition energy in the third row is a consequence of the asymmetric setup 
  which reduces image charge effect from the more distant electrode. In the last
  column the point group symmetry of the Hamiltonian in the absence of an 
  applied bias voltage is given.}
\label{tab:energy_levels}
\end{table}

When taking into account the image charge effect, the symmetry of the full
junction Hamiltonian given by Eqs.~\eqref{eq:H_PPP} and~\eqref{eq:H_PPP_molenv}
is reduced with respect to that of the of the isolated molecule. In this case,
the symmetry of the junction Hamiltonian reflects the symmetry of the combined
molecule plus junction setup. The point groups of the Hamiltonian for the
different setups are listed in the last column of Tab.~\ref{tab:energy_levels}.
For example, in the symmetric setup in Fig.~\ref{fig:benzene_set}b, the symmetry
is reduced to the $D_{2h}$ point group. As a consequence, the orbital degeneracy
of the anion $E_{2u}$ ground-state is lifted resulting in a splitting $\Delta$
between the symmetry-broken states. The situation is illustrated schematically
in the right part of Fig.~\ref{fig:overview}a (the labeling of the states is
explained in the next section). The splitting of the degenerate anion state is
listed in the second column of Tab.~\ref{tab:energy_levels} for the different
setups. In all cases, the splitting is of considerable size ($40-80$ meV). In
the regime $k_\text{B}T > \Gamma$ considered here, the splitting thus exceeds
the level broadening $\Gamma$ even at room-temperature in the Coulomb blockaded
junctions. With the distance to the image plane of the electrodes in the
symmetric setup increased to $\sim 4$ {\AA}, the splitting remains on the order
of 10 meV. Hence, irrespective of the exact alignment between the molecule and
the electrodes, the splitting is large enough for the split states to appear as
individual resonances in the charge stability diagram at sufficiently low
temperatures.

In the junction considered here, the electrodes and the gate dielectric affect
the molecular symmetry differently. Since the molecule is lying flat on the gate
dielectric, it does not break the symmetry of molecule. The image charge effect
from the electrodes is therefore most important for the observed lifting of the
degeneracies. On the other hand, both the gate dielectric and the electrodes
contributes equally to the reduction of the addition
energy~\cite{Kaasbjerg:OPV5SET}.

\subsection{Selection rules}
\label{sec:selection_rules}

The charge transport through the molecule is to a high degree determined by the
transition matrix elements in Eq.~\eqref{eq:gamma}. For molecules with
symmetries, group theoretical arguments can be used to derive selection rules
for the transition matrix elements between the involved states. The selection
rules for the isolated benzene molecule have been considered in detail in
Ref.~\onlinecite{Begemann:InterferenceEffects}. Due to the lower symmetry of the
full junction Hamiltonian considered here, the following analysis differs
slightly.

For the symmetric setup in Fig.~\ref{fig:benzene_set}b the Hamiltonian belongs
to the $D_{2h}$ point group. The elements of the point group and the Hamiltonian
therefore posses a common set of eigenstates. Here, the symmetry of interest is
the symmetry operation $\sigma_v$ which is a reflection about the plane through
the two carbon atoms closest to the electrodes and perpendicular to the
molecular plane. The eigenstates can therefore be classified by the eigenvalues
$\pm 1$ corresponding to symmetric and antisymmetric states with respect to
reflections in the mirror plane $\sigma_v$. The ground-state $\ket{0}$ of the
neutral molecule is symmetric. i.e. $\sigma_v \ket{0} = \ket{0}$. For the split
ground-state of the anion, the lower and higher lying states are symmetric and
antisymmetric, respectively. We therefore label them as in
Fig.~\ref{fig:overview} where $\ket{s}$ denotes the symmetric and $\ket{a}$ the
antisymmetric excited state of the anion.

We now consider coupling in the \emph{para} configuration where the leads couple
to the two atoms facing the electrodes in Fig.~\ref{fig:benzene_set}b. Since the
coupling atoms lie in the mirror plane $\sigma_v$, the transition matrix element
between the symmetric $N=6$ ground-state and the antisymmetric $N=7$ state can
be shown to fulfill the following equality
\begin{equation}
  \label{eq:selection}
  \langle N+1,a \vert  c^{\dagger}_{\alpha\sigma} \vert N,0 \rangle
  = -
  \langle N+1,a \vert c^{\dagger}_{\alpha\sigma} \vert N,0 \rangle ,
\end{equation}
implying that the matrix element vanishes~\cite{Begemann:InterferenceEffects}.
Transitions to the antisymmetric state are therefore forbidden which results in
a so-called dark state that cannot be observed in transport measurements. For
transitions between the two symmetric states no such restriction exist. In other
words, this is a statement that the operator $c^{\dagger}_{\alpha\sigma}$
preserves the symmetry of the state when coupling in \emph{para} configuration.
In $\emph{meta}$ configuration, where one of the couplings is shifted to the
neighboring site, the creation operator for the shifted coupling site no longer
preserves the mirror symmetry of the states. Hence, the matrix element will be
non-zero for both the symmetric and antisymmetric state of the anion.
\begin{figure}[!t]
  \centering
  \includegraphics[width=0.99\linewidth]{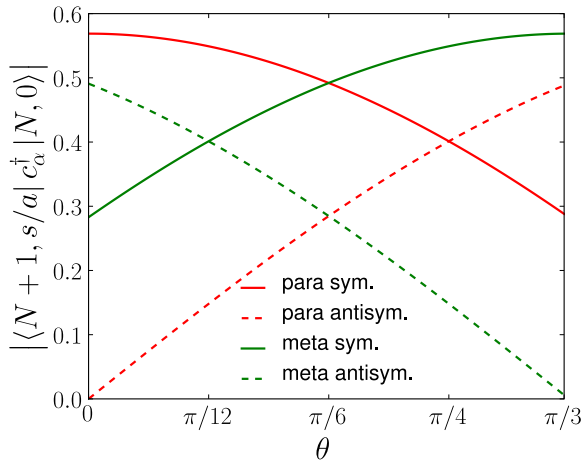}
  \caption{(Color online) Transition matrix elements $\left\vert \langle N+1,s/a
      \vert c^{\dagger}_{\alpha\sigma} \vert N,0 \rangle \right\vert$ between
    the ground-state $\ket{0}$ of the neutral molecule ($N=6$) and the
    symmetric/antisymmetric state $\ket{s}$/$\ket{a}$ of the anion ($N=7$) vs
    angle of rotation of the benzene molecule. The rotation angle is with
    respect to the symmetric setup in Fig.~\ref{fig:benzene_set}b. The red and
    green lines refer to coupling at the \emph{para} and \emph{meta} sites
    denoted by red and green dots in Fig.~\ref{fig:benzene_set}, respectively.}
\label{fig:gamma_vs_theta}
\end{figure}

Since the mirror symmetry $\sigma_v$ is an element of the $C_{2v}$ point group,
the selection rules derived above also apply in the asymmetric setup in
Fig.~\ref{fig:benzene_set}c. One way to break the mirror symmetry, is to rotate
the benzene molecule around its six-fold rotational symmetry axis as sketched in
Fig.~\ref{fig:benzene_set}d. This breaks all the symmetries in the molecule and
destroys the selection rules for the transition matrix elements.
Figure~\ref{fig:gamma_vs_theta} shows the absolute value of the transition
matrix elements as a function of the rotation angle $\theta$ with respect to the
symmetric setup in Fig.~\ref{fig:benzene_set}b. The matrix elements for coupling
to both the \emph{para} (red lines) and \emph{meta} (green lines) site are
shown. As evident from Fig.~\ref{fig:benzene_set}, the \emph{para} coupling
sites become the \emph{meta} site and vice versa under a rotation of
$\theta=\pi/3$. For $\theta=\pi/6$ the two coupling sites are equivalent.
This is reflected in the mirror symmetry between the red and green lines in
Fig.~\ref{fig:gamma_vs_theta} which meet at $\theta=\pi/6$.

As discussed above, the \emph{para} matrix element vanishes for the
antisymmetric state at $\theta=0$. The other transition matrix elements all have
finite values in the non-rotated setup. For $\theta \neq 0$ the mirror symmetry
of the junction is broken and the selection rules derived from
Eq.~\eqref{eq:selection} no longer apply. As a result, the transition matrix
element for the antisymmetric state with coupling at the \emph{para} site
acquires a finite value. Under these circumstances, transport via the otherwise
dark state becomes possible. As will be discussed in further detail below, the
small magnitude of the associated transition matrix element for the
antisymmetric state in Fig.~\ref{fig:gamma_vs_theta} is likely to cause NDR. The
occurrence of NDR from a low-lying symmetry-split state may therefore serve as a
fingerprint of a broken symmetry in the molecule.

\subsection{Current and NDR}

At relatively low biases, only the low-lying states illustrated in
Fig.~\ref{fig:overview}a will be active in the charge transport through the
molecule. More specifically, only the ground-state of the neutral molecule and
the split degenerate ground-state of the singly charged molecule needs to be
considered. Given that the temperature is low enough and $k_\text{B}T > \Gamma$,
the large splittings $\Delta$ of the anion states in
Tab.~\ref{tab:energy_levels} bring the junction outside the quasi-degenerate
regime $\Gamma \sim \Delta$ where coherence between the degenerate states is
important~\cite{Begemann:InterferenceEffects,Schultz:Degeneracies}. Hence, the
out-of-equilibrium occupations of the molecular many-body states and the current
can be obtain with the conventional rate-equation approach described in
Sec.~\ref{sec:rate_equations}.

With only three states---two of which are the symmetry-split doublets of the
singly charged molecule---participating in the transport the stationary
rate-equations take the simple form
\begin{equation}
  \renewcommand{\arraystretch}{1.4}
  \begin{pmatrix} 
    - (\Gamma_{s 0} + \Gamma_{a 0}) & \Gamma_{0 s} & \Gamma_{0 a} \\ 
    \Gamma_{s 0} & - \Gamma_{0 s} & 0 \\ 
    \Gamma_{a 0} & 0 & - \Gamma_{0 a} \\
  \end{pmatrix}
  \begin{pmatrix} 
    P_0 \\ P_{s} \\ P_{a} \\ 
  \end{pmatrix}
  = 0
\end{equation}
where the $0$, $s$ and $a$ subscripts denote the ground-state of the neutral
molecule and the symmetric ground-state and first excited antisymmetric state of
the charged molecule, respectively. The rates $\Gamma_{ij} = \Gamma_{ij}^L +
\Gamma_{ij}^R$ have contributions from tunneling to both the left and right
leads. For the states of the charged molecule $P_{s/a}$ denotes the occupations
of the individual spin up and down states of the doublet which are equal, i.e.
$P_{i\uparrow}=P_{i\downarrow}=P_{i}$. Together with the normalization condition
$P_0 + 2P_{s} + 2P_{a} = 1$ the rate-equations have the solution
\begin{equation}
  P_0 = \frac{\Gamma_{0s}\Gamma_{0a}}{\widetilde{\Gamma}^2} \; ; \;
  P_s = \frac{\Gamma_{s0}\Gamma_{0a}}{\widetilde{\Gamma}^2} \; ; \;
  P_a = \frac{\Gamma_{a0}\Gamma_{0s}}{\widetilde{\Gamma}^2} ,
\end{equation}
where $\widetilde{\Gamma}^2=\Gamma_{0 s}\Gamma_{0 a} + 2( \Gamma_{a0}\Gamma_{0s}
+ \Gamma_{s0}\Gamma_{0a})$. The current through the molecule, here evaluated at
the left lead, is then given by
\begin{align}
  \label{eq:current1}
  I & = - 2e 
  \left[ 
    P_0 (\Gamma_{s 0}^L + \Gamma_{a 0}^L) - 
    P_s \Gamma_{0 s}^L - P_a \Gamma_{0 a}^L
  \right] \nonumber \\
  & = 2e \frac{\Gamma_{s0}^R\Gamma_{0a}\Gamma_{0s}^L + 
               \Gamma_{a0}^R\Gamma_{0s}\Gamma_{0a}^L - 
               \Gamma_{0s}^R\Gamma_{0a} \Gamma_{s0}^L -
               \Gamma_{0a}^R\Gamma_{0s} \Gamma_{a0}^L}
             {\widetilde{\Gamma}^2},
\end{align}
where the factor of 2 comes from the spin degeneracy of the anionic doublet
states.

In the following we wish to establish a general condition for $\gamma$-factors
in Eq.~\eqref{eq:gamma} under which NDR occurs for the three state system of the
symmetry-broken benzene molecule. Similar considerations have been given in
Ref.~\onlinecite{Datta:Generic} for a spinless three state system. We here
generalize this condition to take into account the spin degeneracy of the two
anionic states. The result is completely general and can be applied to similar
three state systems which generally occur in single-molecule junctions when an
excited state of a charged molecule becomes
accessible~\cite{Leijnse:Asymmetric}.

In the case of NDR, the current decreases at the voltage where the excited state
enters the bias window. The situation is illustrated schematically in
Fig.~\ref{fig:overview}c. Here, $I_s$ is the current when only the level for the
symmetric state of the anion is located in the bias window. For larger values of
the bias voltage where the antisymmetric state becomes accessible, the current
either increases or decreases to the value $I_{sa}$. In this case, the current
is given by the expression in Eq.~\eqref{eq:current1}. In the former case, the
Fermi factor in Eq.~\eqref{eq:Gamma_in} is zero for the antisymmetric state and
the expression for the current simplifies to
\begin{align}
  \label{eq:current2}
  I & = -2e
  \left[
    P_0 \Gamma_{s0}^L - P_s \Gamma_{0s}^L 
  \right] \nonumber \\
  & = 2e \frac{\Gamma_{s0}^R\Gamma_{0s}^L - \Gamma_{0s}^R\Gamma_{s0}^L}
            {\Gamma_{0 s} + 2\Gamma_{s0}}  .
\end{align}
The NDR, illustrated by the (red) dashed line in Fig.~\ref{fig:overview}c,
occurs when $\vert I_s \vert > \vert I_{sa} \vert$. At the horizontal plateaus
in between the jumps in the current, the distance $\vert E_{s/a} - \mu_{\alpha}
\vert$ between the molecular levels and the chemical potentials of the leads is
assumed to be larger than the thermal energy $k_\text{B}T$. The Fermi factors
$f_{\alpha}$ and $1-f_{\alpha}$ in Eqs.~\eqref{eq:Gamma_in}
and~\eqref{eq:Gamma_out} for the rates can therefore be set to either unity or
zero. Together with the current expressions in Eqs.~\eqref{eq:current1}
and~\eqref{eq:current2}, the inequality for the currents above leads to the
following conditions for the $\gamma$-factors
\begin{align}
  \label{eq:ndr}
  \frac{1}{\gamma_{0a}^L} > \frac{1}{2\gamma_{s0}^R} + \frac{1}{\gamma_{0s}^L}
  \quad 
  (\mu_L < \mu_R) \nonumber \\
  \frac{1}{\gamma_{0a}^R} > \frac{1}{2\gamma_{s0}^L} + \frac{1}{\gamma_{0s}^R}
  \quad 
  (\mu_L > \mu_R)  ,
\end{align}
which must be fulfilled in order to have NDR at positive or negative bias
voltage, respectively. These inequalities can be fulfilled when the excited
antisymmetric state of the anion is a so-called blocking state which has a small
exit rate, i.e. $\gamma_{0a}^{\alpha}$ must be small for the drain electrode
compared to at least one of the $\gamma$-factors for the symmetric state. The
existence of such a small exit rate results in a slowing down of the charge
transfer dynamics when the corresponding state enters the bias window and is
therefore often accompanied by NDR. In Fig.~\ref{fig:overview}a the
corresponding transition is marked by the (red) dashed arrow. It should be noted
that the inequalities above only require the exit rate for the excited state to
be small and put no constraint on the in rate.

A situation with a small exit rate can be realized in either of the following
two ways (or both). In the case of strongly asymmetric tunnel couplings, e.g.
$t_L \ll t_R$, the exit rate becomes small when the coupling to the drain
electrode is the weaker. However, given that the transition matrix elements for
the states in Eq.~\eqref{eq:gamma} are the same, this situation results in a
small exit rate for both the symmetric and antisymmetric state and is therefore
not sufficient for the occurrence of NDR. A difference in the transition matrix
elements between the states in Eq.~\eqref{eq:gamma} on top of the asymmetry in
the tunnel couplings is therefore required to fulfill the inequalities in
Eq.~\eqref{eq:ndr}. On the other hand, for symmetric tunnel couplings, i.e. $t_L
= t_R$, NDR is also possible. Here, the inequalities Eq.~\eqref{eq:ndr} may be
satisfied by the transition matrix elements alone when the matrix element for
the excited state is small. A recent theoretical study has demonstrated that
this property can be designed into conjugated molecules by functionalizing them
with different types of chemical groups~\cite{Leijnse:Asymmetric}. This results
in asymmetric molecular orbitals for the excited states and consequently a small
transition matrix element at one end of the molecule that produces NDR. As
discussed in the previous section, the small transition matrix element is here
provided by the destroyed selection rules in the low-symmetry setup in
Fig.~\ref{fig:benzene_set}d where the molecule is placed in a rotated
configuration.

\section{IV characteristics and stability diagrams}
\label{sec:iv}

\begin{figure*}[!t]
  \centering
  \begin{minipage}{0.49\textwidth}
    \includegraphics[width=0.9\linewidth]{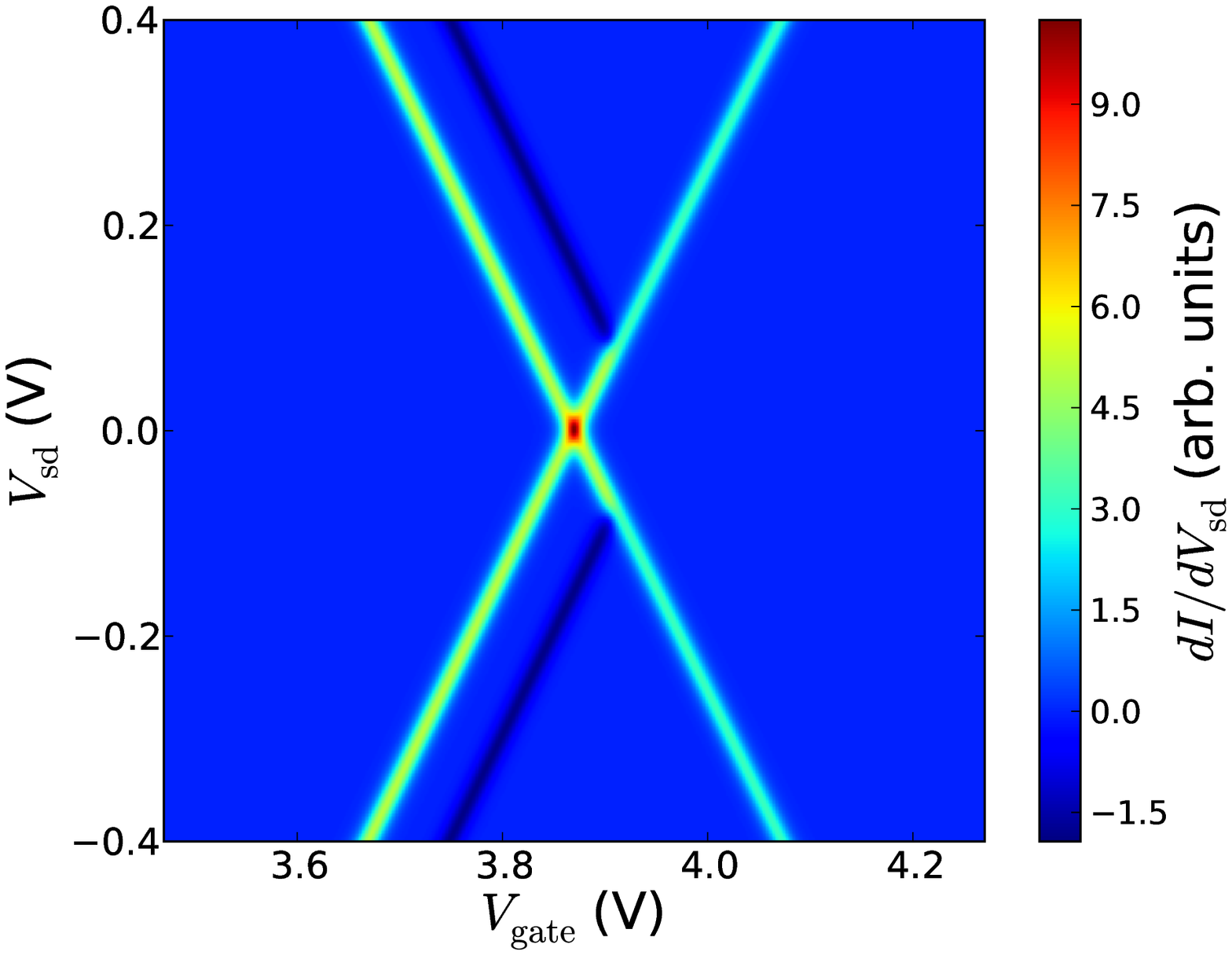}
  \end{minipage} \hfill
  \begin{minipage}{0.49\textwidth}
    \includegraphics[width=0.9\linewidth]{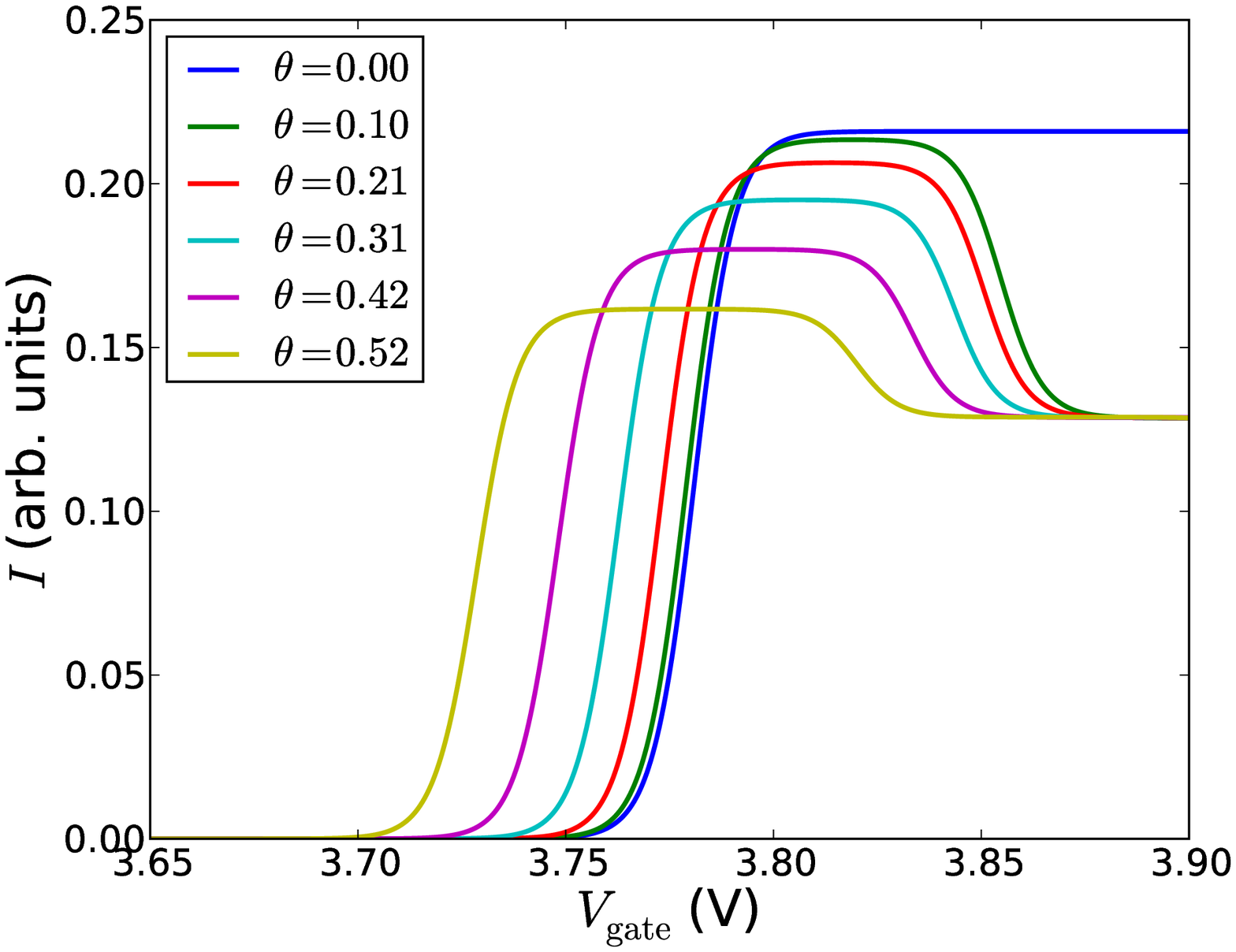}
  \end{minipage}
  \caption{(Color online) IV characteristic for the symmetric setup and coupling
    in the \emph{para} configuration. Left: Stability diagram for
    $\theta=\pi/12\sim0.26$, $k_\text{B}T=5$ meV and $t_L=t_R$. Right: Absolute
    value of the current vs gate voltage at $V_{\text{sd}}=0.2$ V for different
    values of $\theta$. As evident from the blue $\theta=0$ curve, the selection
    rules for the transition matrix elements in Eq.~\eqref{eq:gamma} completely
    expel the antisymmetric anionic state from participating in the transport
    and no NDR is observed. The varying position of the current onset and the
    current dip is caused by a weak $\theta$-dependence of the molecular energy
    levels.}
\label{fig:stability_sym}
\end{figure*}
In the following section the IV characteristics of the different setups is
considered. In all cases the temperature of the lead electrons is set to
$k_\text{B}T=5$~meV and the equilibrium Fermi level of the leads is positioned
in the middle of the molecular gap. Furthermore, since only transport via the
negatively charged anion of the molecule is considered, the discussion is
restricted to positive gate voltages. It should be emphasized that the many-body
states, their energy, and the transition matrix elements in Eq.~\eqref{eq:gamma}
are calculated for each value of the applied voltages. The IV characteristics
reported in the following therefore include possible shifts of the molecular
levels with the applied bias voltage. In particular, different capacitive
couplings to the source and drain electrodes arising from the asymmetric setup
will result in IV characteristics that are asymmetric in the bias voltage.

\subsection{Symmetric setup with coupling in the \emph{para} configuration}

Figure~\ref{fig:stability_sym} summarizes the IV characteristics of the
symmetric setup in Fig.~\ref{fig:benzene_set}b. Due to the selection rules, the
antisymmetric state is not active in the transport for the non-rotated case when
coupled in the \emph{para} configuration. However, when the molecule is rotated
the matrix element for the antisymmetric state acquires a finite value (see
Fig.~\ref{fig:gamma_vs_theta}) allowing the antisymmetric state to be populated
when it enters the bias window. The left plot in Fig.~\ref{fig:stability_sym}
shows the stability diagram for a rotation angle $\theta=\pi/12$ of the benzene
molecule and symmetric tunnel couplings $t_L=t_R$. At this value for $\theta$,
the ratio between the transition matrix elements for the symmetric and
antisymmetric states with coupling to the \emph{para} site is $\left\vert
  \langle s \vert c^{\dagger}_{\alpha\sigma} \vert 0 \rangle / \langle a \vert
  c^{\dagger}_{\alpha\sigma} \vert 0 \rangle \right\vert \sim 3.7$. Reverting to
the condition for the occurrence of NDR in Eq.~\eqref{eq:ndr}, this value for
the ratio between the transition matrix elements is found to fulfill the
inequalities in the case of symmetric tunnel couplings. Consequently, an NDR
feature appears in the stability diagram at the voltages where the antisymmetric
state becomes accessible. The right plot in Fig.~\ref{fig:stability_sym} shows
the absolute value of the current as a function of the gate voltage at
$V_{\text{sd}}=0.2$ V for different values of the rotation angle of the benzene
molecule. Due to the symmetry forbidden population of the antisymmetric state at
$\theta=0$, this state remains unpopulated at the gate voltage where it enters
the bias window. Therefore, the corresponding curve in the right plot of
Fig.~\ref{fig:stability_sym} does not show any change in the current at the
corresponding gate voltage. For increasing rotation angles, both the NDR effect
and the current level prior to the onset of the NDR is seen to decrease. This
trend is a direct consequence of the variation of the transition matrix elements
shown in Fig.~\ref{fig:gamma_vs_theta} when the molecule is rotated with respect
to the electrodes. Likewise, the shift of the onset for the current and NDR
originates from small changes in the level positions when the molecule is
rotated.

The linear character of the Coulomb diamond boundaries reveals that the bias
voltage does not have any apparent effect on the molecular states and their
energies. Only at much higher voltages may this become important. For longer
molecules, however, where the voltage drop over the molecule becomes
significant, a larger impact on the molecular states is
possible~\cite{Kaasbjerg:OPV5SET}. These consideration also indicate that the
symmetry-breaking effect of the bias voltage is here negligible compared to that
of the image charge effect.

\subsection{Asymmetric setup with coupling in both \emph{para} and \emph{meta}
  configuration}

\begin{figure*}[!t]
  \centering
  \begin{minipage}{0.49\textwidth}
    \includegraphics[width=0.9\linewidth]{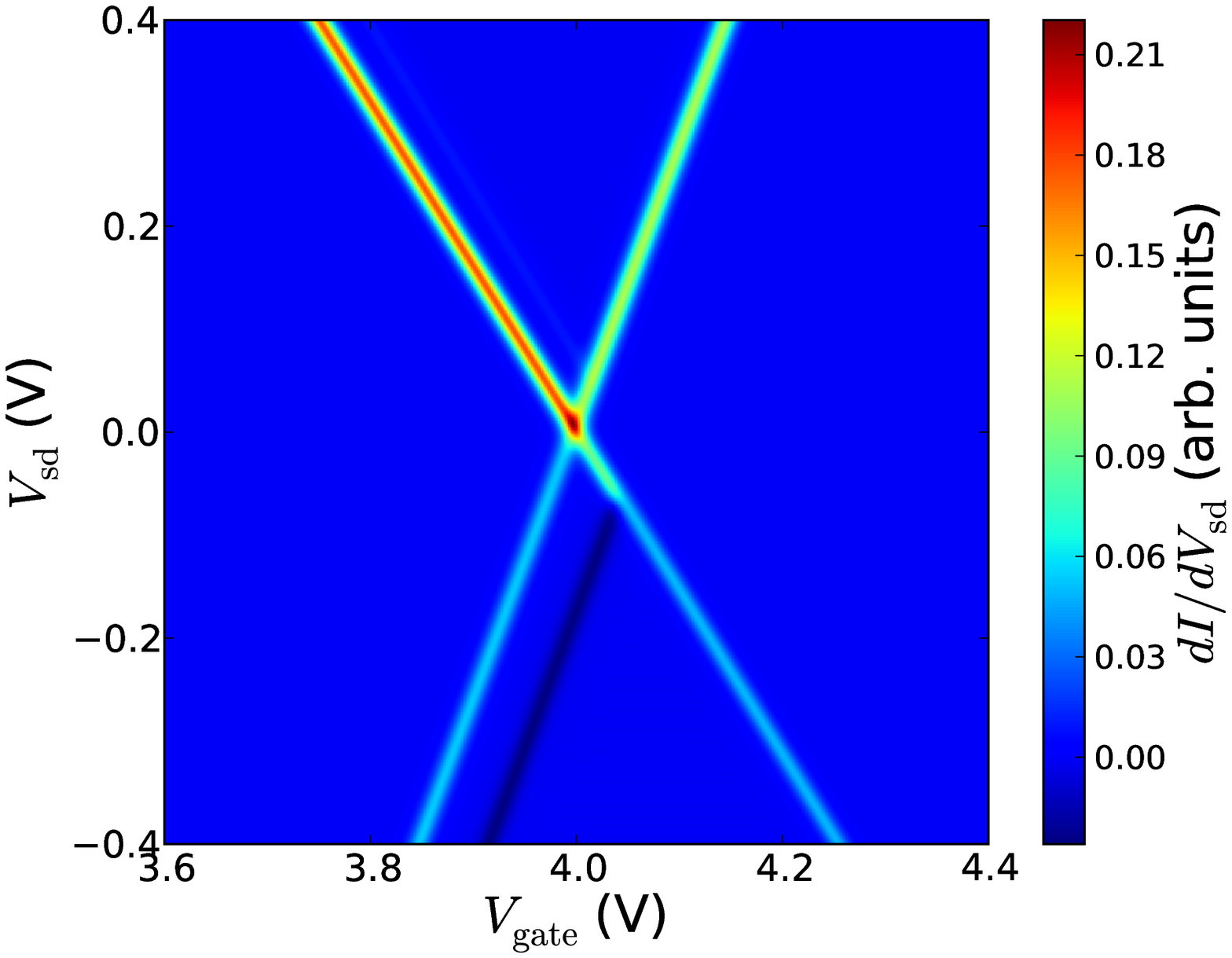}
  \end{minipage} \hfill
  \begin{minipage}{0.49\textwidth}
    \includegraphics[width=0.9\linewidth]{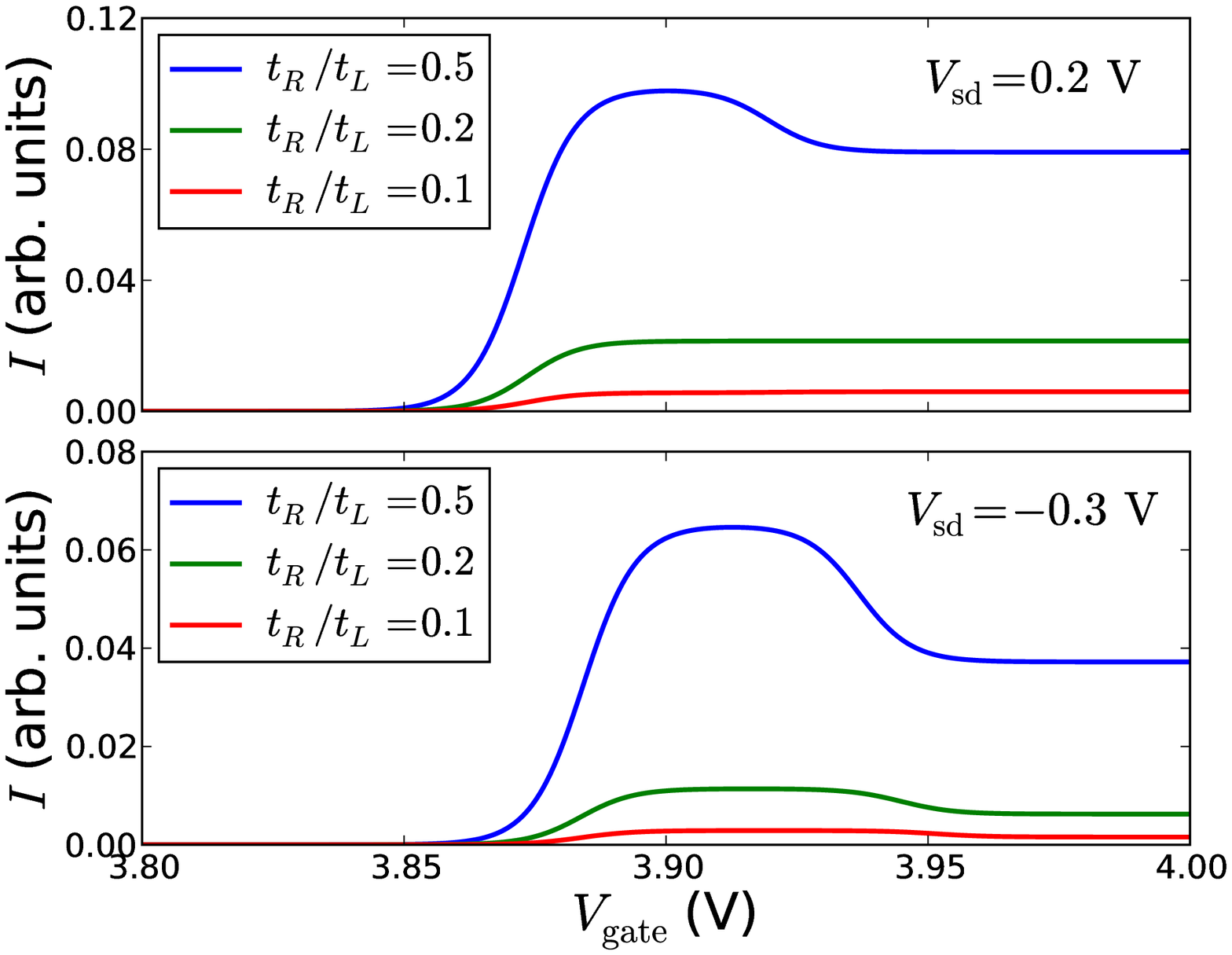}
  \end{minipage}
  \caption{(Color online) IV characteristics for the asymmetric setup where the
    distance to the right electrode is twice the distance to the left electrode
    and with coupling in the \emph{para} configuration. Left: Stability diagram
    for $\theta=\pi/12\sim0.26$, $k_\text{B}T=5$ meV and $t_L=10t_R$. Right:
    Absolute value of the current vs gate voltage for different values of the
    tunnel coupling ratio $t_R/t_L$ at positive and negative bias voltage.}
\label{fig:stability_asym_para}
\end{figure*}
Next, we focus on the asymmetric setups illustrated in
Fig.~\ref{fig:benzene_set}c. Due to the larger distance to the right electrode
it becomes relevant to address the effect of an asymmetry in the tunnel
couplings, i.e. $t_L > t_R$, and coupling to the \emph{meta} site at the right
electrode. Since the values of the transition matrix elements in
Fig.~\ref{fig:gamma_vs_theta} are left relatively unaffected by the asymmetry of
the setup, changes in the IV characteristics as compared to the symmetric setup
considered in the previous section, can be attributed to the introduced tunnel
coupling asymmetry and the change in the coupling site.

\subsubsection{Coupling in \emph{para} configuration}

Figure~\ref{fig:stability_asym_para} summarizes the IV characteristics for the
asymmetric setup with coupling in the \emph{para} configuration. Here, the left
plot shows the stability diagram for $\theta=\pi/12$ and $t_R/t_L=0.1$. Due to
the asymmetry of the setup, the different capacitive couplings to
the left and right electrodes give rise to different slopes of the
diamond edges in the stability diagram. 

A clear change in the stability diagram compared to the one for the symmetric
setup in Fig.~\ref{fig:stability_sym}, is the disappearance of the NDR feature
for positive bias voltages. This behavior follows from the asymmetry in the
tunnel couplings. For positive bias voltages the electrons exit the molecule to
the left electrode. Depending on the ratio between the transition matrix
elements, the asymmetry in the tunnel couplings may result
in a large exit rate, i.e. $\gamma_L > \gamma_R$. From the NDR conditions in
Eq.~\eqref{eq:ndr}, one sees that this can remove the NDR. Indeed, for
$t_R/t_L=0.1$ and the previous stated ratio of the transition matrix elements,
this is the case. For these parameter values, the difference in the transition
matrix elements for the symmetric and antisymmetric states that produced NDR in
the case of symmetric couplings, is outweighed by the large asymmetry in the
tunnel couplings. Hence, the NDR feature is absent at positive bias voltages. On
the other hand, for negative bias voltage where the NDR feature still appears,
the exit rate for the antisymmetric state to the right electrode remains small
and the lower inequality in Eq.~\eqref{eq:ndr} is fulfilled.

Since the tunnel couplings depend exponentially on the distance to the
electrodes, other values for their ratio should also be considered. The right
plot in Fig.~\ref{fig:stability_asym_para} shows current as a function of gate
voltage for different values of the ratio $t_R/t_L$ at positive ($V_{\text{sd}}
= + 0.2$ V) and negative ($V_{\text{sd}} = - 0.3$ V) bias voltage. As expected,
the current level and the dip in the current occurring at the position of the
NDR feature in the stability diagram, are highly dependent on the tunnel
couplings. At positive bias, the current dip appears only for the smallest
asymmetry $t_R/t_L=0.5$ in the couplings. From the ratio between the transition
matrix elements for the symmetric and antisymmetric state given in the previous
section, the value of the ratio $t_R/t_L$ at which no dip in the current is
observed can be deduced from Eq.~\eqref{eq:ndr}. For positive bias, the upper
inequality in Eq.~\eqref{eq:ndr} becomes an equality at $t_R/t_L \sim 0.2$. At
this value for the coupling ratio, the difference in the transition matrix
elements has been compensated for by the asymmetry in the tunnel couplings and
no change in the current occurs when the antisymmetric state enters the bias
window. For smaller values of the coupling ratio, i.e. larger asymmetry in the
tunnel couplings, the current increases slightly. Again, at negative bias
voltage, the tunnel coupling to the right drain electrode is the smaller and the
current dip persists for all the shown values of the coupling ratio.

\subsubsection{Coupling in \emph{meta} configuration}

\begin{figure*}[!t]
  \centering
  \begin{minipage}{0.49\textwidth}
    \includegraphics[width=0.9\linewidth]{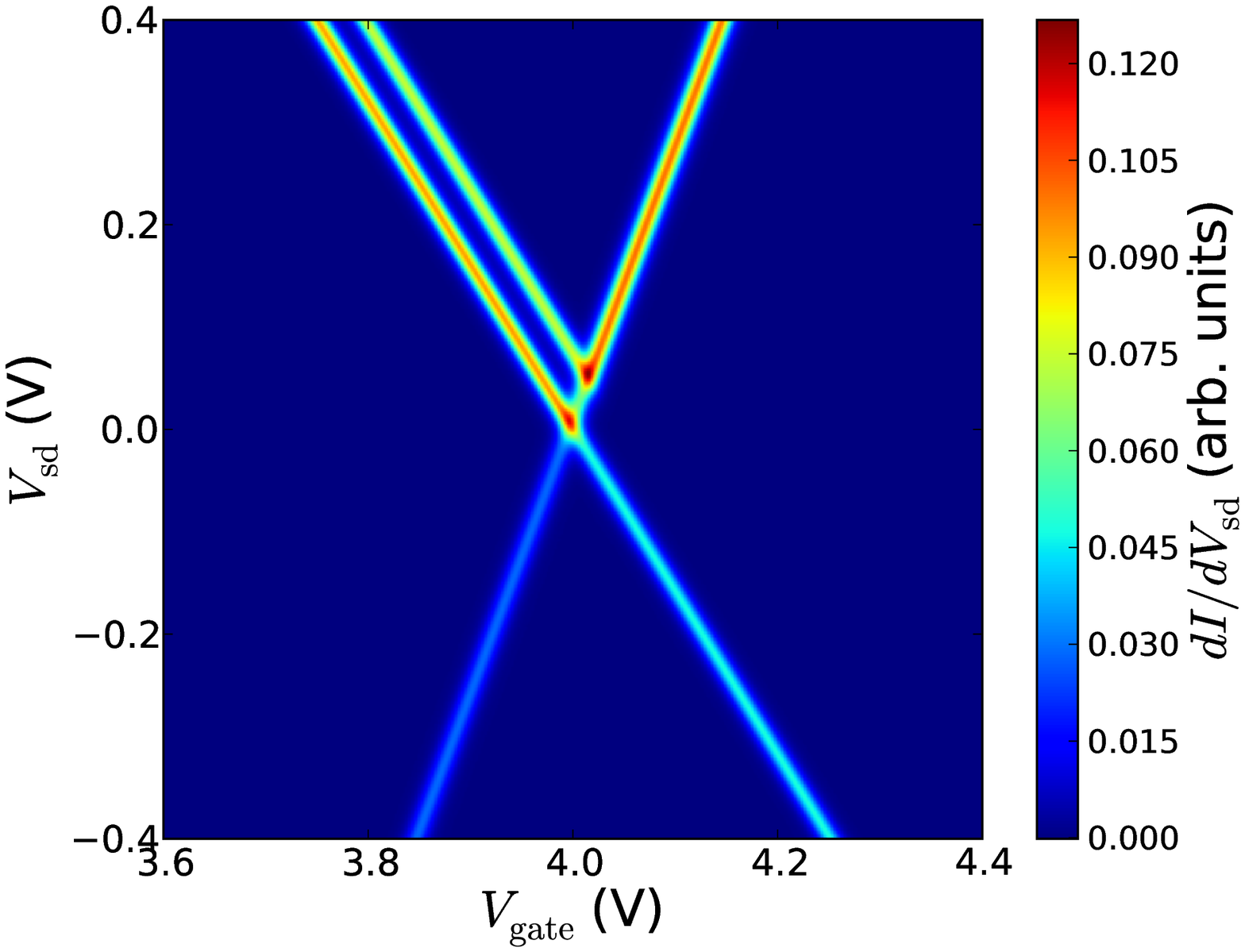}
  \end{minipage} \hfill
  \begin{minipage}{0.49\textwidth}
    \includegraphics[width=0.9\linewidth]{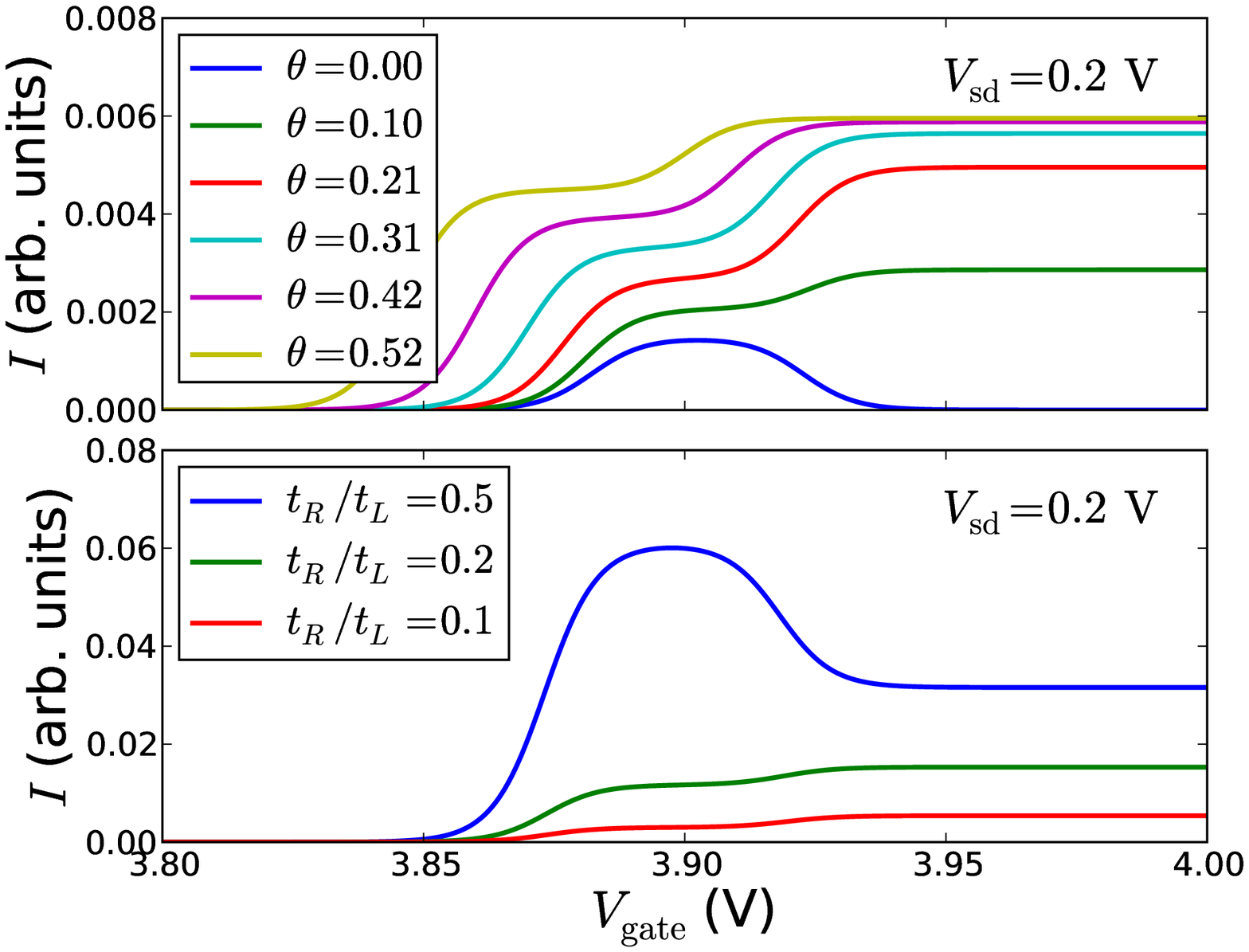}
  \end{minipage} \hfill
  \caption{(Color online) IV characteristics for the asymmetric setup with
    coupling in the \emph{meta} configuration. Left: Stability diagram for the
    same set of parameters as in Fig.~\ref{fig:stability_asym_para}. Right:
    Absolute value of the current vs gate voltage for different values of the
    rotation angle $\theta$ at $t_L=10t_R$ (upper) and the tunnel coupling
    ratio $t_R/t_L$ at $\theta=\pi/12$ (lower). Note the collapse of the current
    for $\theta=0$ in the upper plot.}
\label{fig:stability_asym_meta}
\end{figure*}
Changing the coupling site at the right electrode to the \emph{meta} site
changes things drastically. Figure~\ref{fig:stability_asym_meta} summarizes the
IV characteristics for this case. Again, the left plot shows the stability
diagram for the same parameter values as used in
Fig.~\ref{fig:stability_asym_para}. As shown in Fig.~\ref{fig:gamma_vs_theta},
the transition matrix elements for coupling to the \emph{meta} site are equal at
$\theta=\pi/12$ for the symmetric and antisymmetric state. This has the
immediate consequence that $\gamma_{0a}^R = \gamma_{0s}^R$, implying that the
lower inequality in Eq.~\eqref{eq:ndr} giving the condition for NDR at negative
bias can never be fulfilled. This is reflected in the stability diagram where no
NDR from the antisymmetric state is observed at negative bias. As opposed to the
situation with coupling in the \emph{para} configuration, the resonance from the
antisymmetric state now appears at positive bias and gives rise to an increase
in the current instead of NDR.

The upper right plot in Fig.~\ref{fig:stability_asym_meta} shows how the current
in the vicinity of this resonance varies with the rotation angle $\theta$. For
$\theta=0$ a complete collapse of the current is observed when the antisymmetric
state becomes accessible. From the selection rules discussed in
Sec.~\ref{sec:selection_rules} and Fig.~\ref{fig:gamma_vs_theta}, we recall that
the transition matrix element for coupling to the \emph{para} site vanishes,
while the matrix element for coupling to the \emph{meta} site is finite. At
positive bias this allows the antisymmetric state to become populated from the
right electrode. However, since the exit process from the \emph{para} site at
the left electrode is forbidden by symmetry, the molecule remains trapped in the
antisymmetric state and the current is blocked. A similar collapse of the
current in a benzene SET has been reported in
Ref.~\onlinecite{Hettler:CurrentCollapse} where the blocking state was populated
via radiative relaxation from a higher lying excited state. For $\theta \neq 0$
the current collapse as well as the NDR disappears, and an increase in the
current as the one in the stability diagram appears instead.

As in the previous section, the disappearance of the NDR feature in the
stability diagram is caused by the large asymmetry in the tunnel couplings. The
ratio of the couplings at which the current dip vanishes, can be inferred from
the upper inequality in Eq.~\eqref{eq:ndr}. For $\theta=\pi/12$, we find that
the condition for NDR is met for $t_R/t_L > 0.41$. The lower right plot in
Fig.~\ref{fig:stability_asym_meta} shows the current as a function of gate
voltage at $V_{\text{sd}}=+0.2$ for different values of the tunnel couplings. As
expected, only the upper blue curve with $t_R/t_L=0.5$ shows NDR.

\section{Summary and conclusions}
\label{sec:conclusion}

Using a theoretical framework developed for semiconductor nanostructures, we
have studied the impact of the image charge effect on the molecular states and
the transport in a benzene single-electron transistor operating in the Coulomb
blockade regime. As demonstrated, the image charge effect renormalizes the
charging energies and lifts the degeneracy of the twofold orbitally degenerate
ground-state of the singly-charged anion of the benzene molecule. With the
resulting splitting of the degenerate states exceeding both the thermal energy
$k_\text{B}T$ and the level broadening $\Gamma$ from the tunnel coupling to the
leads, this has important consequences for the low-bias IV characteristics of
the benzene SET. In particular, the selections rules between the transport
active many-body states of the molecule are destroyed, which gives rise to the
appearance of a blocking state that leads to the occurrence of NDR. From the
derived NDR conditions for a generic three-state system, the coming and going of
the NDR feature at different parameter values has been analyzed. It is
demonstrated that the appearance of the NDR feature is very sensitive to
asymmetries in the tunnel couplings to the source and drain electrodes, to the
bias polarity, and to changes in the coupling from the \emph{para} site to the
\emph{meta} site. In experimental situations, observations of the described
transport characteristics may be an indication of a broken symmetry in the
molecule.

Altogether, we have demonstrated that image charge effects play a potentially
important role, not only for the position of the molecular levels, but also for
the molecular states and their degeneracies. As mentioned in the introduction,
experimental studies~\cite{Zant:OPV5Excitations,Balestro:NatC60Cotunneling} have
already speculated that image charge effects affect spin-excitations of
molecules. With the exchange coupling in effective spin Hamiltonians being
determined by hopping integrals and Coulomb matrix elements between the
orbitals, a large impact on spin states could be anticipated.

With the present work, we have paved the way for more detailed descriptions of
single-molecule SETs taking into account image charge effects. In this respect,
the electrostatic Green's function for the generic junction geometry given in
App.~\ref{app:SimpleJunction} and the inclusion of the additional image charge
terms in the PPP Hamiltonian in App.~\ref{app:MatrixElements} provide a good
starting point for future studies of the image charge effect in single-molecule
junctions.

\begin{acknowledgments}
  The authors would like to thank M. Leijnse for useful comments on the
  manuscript. The research leading to these results has received funding from
  the European Union Seventh Framework Programme (FP7/2007-2013) under grant
  agreement n$^\circ$ 270369. KK has been partially supported by the Center on
  Nanostructuring for Efficient Energy Conversion (CNEEC) at Stanford
  University, an Energy Frontier Research Center funded by the U.S. Department
  of Energy, Office of Science, Office of Basic Energy Sciences under Award
  Number DE-SC0001060.
\end{acknowledgments}

\appendix

\section{Electrostatic Green's function for a generic junction geometry}
\label{app:SimpleJunction}

\begin{figure}[!t]
  \centering
  \includegraphics[width=0.98\linewidth]{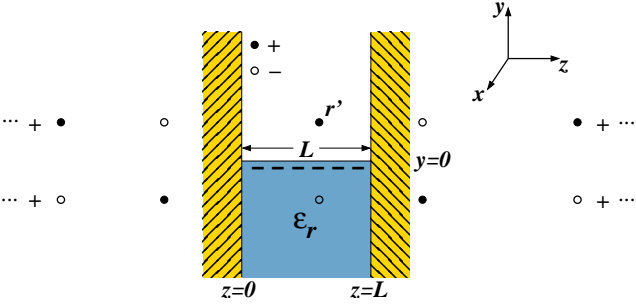}
  \caption{(Color online) Simple junction geometry for which Poisson's equation
    can be solved analytically for the electrostatic Green's function. The
    junction consists of two infinite parallel electrode plates where the lower
    half-space ($y<0$) between them is filled with an oxide of dielectric
    constant $\varepsilon_r$. The dots illustrates the image charge solution to
    Poisson's equation given in Eq.~\eqref{eq:analytic}.}
\label{fig:simple_junction}
\end{figure}

In this appendix we give an analytical solution of Poisson's equation for the
electrostatic Green's function in the simplified junction geometry shown in
Fig.~\ref{fig:simple_junction}. Despite its simple structure, the Green's
function of this junction provides a good description of the potential in more
realistic junctions as the one illustrated in Fig.~\ref{fig:benzene_set}.

When solving Poisson's equation in Eq.~\eqref{eq:PoissonGF}, the screening
induced polarization charge of dielectric regions is most often accounted for by
the spatially dependent dielectric constant $\varepsilon_r(\br)$. In the
following a different route will be taken by treating the polarization charge as
a source term in Poisson's equation. For the homogeneous gate dielectric in
Fig.~\ref{fig:simple_junction}, the polarization charge induced by a unit source
charge in $\br'$ will be a surface charge $\sigma$ that resides on the interface
between the dielectric and the vacuum region. The Green's function can therefore
be obtained from the following Poisson equation
\begin{equation}
  \label{eq:AnalyticGF1}
  - \epsilon_0 \nabla^2 G(\br,\br') = 
              \delta(\br-\br') + \frac{\sigma(\br)}{e} 
\end{equation}
with Dirichlet boundary conditions, i.e. $G=0$ at the surface of the metallic
electrodes. The surface charge on the right-hand side is located in the
$xz$-plane $\sigma(\br) = \sigma(x,z)\delta(y)$ and can be obtained as the
normal component of the polarization $\mathbf{P}$ at the interface. Using the
relations $\mathbf{D}=\epsilon_0\mathbf{E} + \mathbf{P}$ and
$\mathbf{D}=\epsilon_0\varepsilon_\text{r} \mathbf{E}$ between the displacement
field $\mathbf{D}$ and the electric field $\mathbf{E}$ in linear dielectrics,
the surface charge can be related to the electric field via
\begin{equation}
  \sigma = \mathbf{P} \cdot \hat{\mathbf{n}} 
         = \epsilon_0 \left(\varepsilon_\text{r} -1\right) \mathbf{E} \cdot \hat{\mathbf{n}} ,
\end{equation}
where the normal component (in this case the $y$-component) is evaluated
immediately below the interface, i.e. $y=0^-$. Expressing the electric field by
the gradient of the Green's function, the following relation between the Green's
function and the surface charge is obtained
\begin{equation}
  \label{eq:AnalyticGF2}
  - \hat{\mathbf{n}} \cdot \nabla G(\br,\br')  
  = \frac{1}{\epsilon_0 (\varepsilon_r-1) } \sigma(\br) .
\end{equation}
This relation can be used to eliminate the surface charge in
Eq.~\eqref{eq:AnalyticGF1}. The resulting equation for the Green's function is
solved by expanding in plane-waves and sines as
\begin{equation}
  G(\br, \br') = \int \frac{dp}{2\pi} \int \frac{dk}{2\pi} \sum_n 
      \sin{\frac{n\pi z}{L}} e^{iky} e^{ipx} G(nkp, \br') .
\end{equation}
After much algebra, the following solution for the electrostatic Green's
function is found
\begin{widetext}
\begin{align}
  \label{eq:analytic}
   G(\br,\br') = &
      \frac{1}{4\pi\epsilon_0} \sum_{\sigma=\pm 1} \sum_{\tau=\pm 1} 
         \sigma\tau \left(\frac{\varepsilon_r+\tau}  {\varepsilon_r+1}\right)
         \times 
        \Bigg[ \frac{1}{\sqrt{(x-x')^2 + (y - \tau y')^2 + (z - \sigma z')^2}} + \nonumber \\
    &    \sum_{n=1}^{\infty} 
         \bigg( 
              \frac{1}{\sqrt{(x-x')^2 + (y-\tau y')^2 + (2nL - (z-\sigma z')^2)}} +
              \frac{1}{\sqrt{(x - x')^2 + (y-\tau y')^2 + (2nL + (z-\sigma z'))^2}}
         \bigg) 
        \Bigg] \nonumber \\
    \equiv & \frac{1}{\abs{\br - \br'}} + \widetilde{G}(\br, \br')  ,
\end{align}
\end{widetext}
where $\mathbf{r},\mathbf{r}'$ belong to the vacuum region of the junction. The
solution has the intuitive image charge interpretation illustrated in
Fig.~\ref{fig:simple_junction}. The analytic solution in Eq.~\eqref{eq:analytic}
allows for a direct identification of the two contributions to the Green's
function indicated in the last equality. Here, the direct Coulomb interaction is
given by the first term inside the square brackets for $\sigma=\tau=+1$, while
the remaining terms give the induced potential $\widetilde{G}$. By comparing to
finite element solutions of Poisson's equation~\eqref{eq:PoissonGF} in junction
geometries as the one illustrated in Fig.~\ref{fig:benzene_set}, we found that
the Green's function for the simplified junction considered here to a high
degree resembles that of more realistic junctions.

As mentioned in the main text the distances between the atoms of the molecule
and the electrostatic boundaries of the junction must be chosen with care. The
reason for this is that the positions of the electrostatic boundaries between
metallic/dielectric regions and the vacuum region where the molecule resides do
not correspond to the actual positions of the atomic surfaces in the junction.
Typically, this so-called electrostatic image plane of the atomic surfaces lie
$\sim 1$ {\AA} outside the outermost atomic layer~\cite{Needs:Image}. The distance
between the atoms of the molecule and the surface atoms are therefore larger
than the chosen distance to the respective image planes.

\section{Image charge Hamiltonian in a localized basis}
\label{app:MatrixElements}

In this appendix the derivation of the PPP representation in
Eq.~\eqref{eq:H_PPP_molenv} of the image charge related terms in the junction
Hamiltonian in Eq.~\eqref{eq:H_molenv} is outlined.

Within the PPP description, the Hamiltonian is expressed in the basis of the
p$_z$-orbitals on the carbon atoms. In this basis the terms related to the image
charge effect in Eq.~\eqref{eq:H_PPP_molenv} read
\begin{equation}
  \label{eq:H_PPP_molenv_full}
  H_{\text{mol-env}} + H_{\text{env}} = \sum_{ij,\sigma} 
      \widetilde{V}_{ij}^{\text{ion}} c^{\dagger}_{i\sigma} c^{\phantom\dagger}_{j\sigma}
      + \frac{1}{2} \sum_{ij, \sigma \sigma'} n_{i\sigma} \widetilde{V}_{ij}
      n_{j\sigma'} .
\end{equation}
Here, $\widetilde{V}_{ij}^{\text{ion}}$ $(>0)$ and $\widetilde{V}_{ij}$ $(<0)$
are the matrix elements of the image charge potential from the ionic
cores and the two-electron integrals of the image charge interaction between the
electrons. The two types of matrix elements are given by
\begin{equation}
  \label{eq:V_tilde_ion}
  \widetilde{V}_{ij}^{\text{ion}} = \int \! d\br \;\phi^*_i(\br) 
     \widetilde{V}_{\text{ion}}(\br) \phi_j(\br) 
\end{equation}
and
\begin{equation}
  \label{eq:V_tilde_image}
  \widetilde{V}_{ij} = \int \! d\br \! \int \! d\br'
  \abs{\phi_i(\br)}^2 \widetilde{G}(\br,\br')
  \abs{\phi_j(\br')}^2 ,
\end{equation}
respectively, where the induced potential from the ions is given by
\begin{align}
  \widetilde{V}_{\text{ion}}(\br) & = - \int \! d\br' \widetilde{G}(\br, \br')
      \rho_{\text{ion}}(\br') \nonumber \\
  & \approx - \sum_i \widetilde{G}(\br, \mathbf{R}_i) .
\end{align}
In the last equality, the charge distribution of the ionic cores have been
approximated by $\delta$-functions located at the positions $\mathbf{R}_i$ of
the carbon atoms in the molecule. Notice that the image potential from the
positively charged ions lifts the onsite energy of the electrons. The matrix
elements of the image charge interaction between the electrons in
Eq.~\eqref{eq:V_tilde_image} are negative and therefore correspond to a
screening of the direct Coulomb interactions between the electrons.

Here we adopt the level of simplicity of the PPP Hamiltonian and neglect
off-diagonal matrix elements of the first term in
Eq.~\eqref{eq:H_PPP_molenv_full} and include only the direct matrix elements of
the image charge interaction in the second term. This results in the Hamiltonian
in Eq.~\eqref{eq:H_PPP_molenv} with the additional terms from the applied
voltages treated similarly.

The matrix elements in Eqs.~\eqref{eq:V_tilde_ion} and~\eqref{eq:V_tilde_image}
are evaluated by approximating the absolute square of the orbitals by
$\delta$-functions centered at the atomic sites,
\begin{equation}
  \widetilde{V}_i^{\text{ion}} = \int \! d\br \; \abs{\phi_i(\br)}^2 
      \widetilde{V}_{\text{ion}}(\br) 
  \approx \widetilde{V}_{\text{ion}}(\mathbf{R}_i)
\end{equation}
and
\begin{equation}
  \widetilde{V}_{ij} = \int \! d\br \int \! d\br' \;
      \abs{\phi_i(\br)}^2 \widetilde{G}(\br, \br') \abs{\phi_j(\br')}^2 
  \approx \widetilde{G}(\mathbf{R}_i, \mathbf{R}_j) ,
\end{equation}
respectively. The matrix elements $\widetilde{V}_{ij}$ obtained for the junction
geometry described in the main text, are shown in
Fig.~\ref{fig:matrix_elements}. Also shown are the matrix elements of the direct
Coulomb interaction given by the Ohno parametrization in Eq.~\eqref{eq:Ohno}. As
the figure shows, the matrix elements of the image charge interaction leads to
significant screening of the interactions on the molecule.

\bibliography{journalabbreviations,references}

\begin{thebibliography}{40}
\bibitem[{not()}]{note1}
\bibinfo{note}{The bond lengths in benzene molecule are chosen to
  $r_{_{\text{C-C}}}=1.397$~{\AA} and $r_{_{\text{C-H}}}=1.1$~{\AA},
  respectively. As only the p$_z$ orbitals on the carbon atoms are treated
  explicitly in the diagonalization of the Hamiltonian, the length
  $r_{_{\text{C-H}}}$ of the bond between C and H only matters for the
  alignment of the molecule with respect to the electrodes.}
\end{thebibliography}

\end{document}